\documentclass[12pt]{iopart}
\makeatletter
\@addtoreset{equation}{section}
\makeatother
\usepackage{setstack,iopams,citesort,subeqn}
\def\d{{\rm d}}

\def\Mmae{\hspace*{-1em}}

\def\Xmae{\hspace*{-1ex}}

\def\Musr{\hspace*{1em}}
\def\Xusr{\hspace*{1ex}}
\def\DMusr{\hspace*{2em}}

\def\Define{:=}

\def\dominance{\mathop{\stackrel{\rm d}{<}}}

\def\nnn{\nonumber\\}
\def\nn{\nonumber}
\def\Check{\check}
\def\text{\mbox}
\def\CMP{{\it Commun. Math. Phys.}\ }
\def\End{{\rm End}}

\def\QED{\hfill$\square$}
\newcommand{\dfrac}[2]{\displaystyle\frac{#1}{#2}}

\newcommand{\Proof}{{\bf Proof. }}

\newtheorem{theorem}{Theorem}[section]

\newtheorem{proposition}[theorem]{Proposition}
\newtheorem{lemma}[theorem]{Lemma}

\begin{document}
\hfill{\tt math-ph/0105049}

\title[The $A_{N-1}$- and $B_N$-Calogero models]%
{%
  Algebraic study on the $A_{N-1}$- and $B_N$-Calogero models with
  bosonic, fermionic and distinguishable particles
}
\author{Akinori Nishino\dag \ and Hideaki Ujino\ddag}

\address{\dag \ %
  Department of Physics, Graduate School of Science,
  University of Tokyo,\\
  Hongo 7--3--1, Bunkyo-ku, Tokyo 113--0033, Japan
}
\address{\ddag \ %
  Gunma College of Technology,\\
  Toriba-machi 580, Maebashi-shi, Gunma-ken 371--8530, Japan
}

\address{E-mail: nishino@monet.phys.s.u-tokyo.ac.jp and %
ujino@monet.phys.s.u-tokyo.ac.jp}

\begin{abstract}
  Through an algebraic method using the Dunkl--Cherednik operators,
  the multivariable Hermite and Laguerre polynomials associated
  with the $A_{N-1}$- and $B_N$-Calogero models with bosonic, fermionic and
  distinguishable particles are investigated. The Rodrigues formulas of
  column type that algebraically generate
  the monic non-symmetric multivariable Hermite and Laguerre polynomials
  corresponding to the distinguishable case are presented. Symmetric and
  anti-symmetric polynomials that respectively give the eigenstates
  for bosonic and fermionic particles are also presented by the
  symmetrization and anti-symmetrization of the non-symmetric ones.
  The norms of all the eigenstates for all cases are algebraically
  calculated in a unified way.
\end{abstract}

\footnotetext{Published in J. Phys. A: Math. Gen. {\bf 34} (2001) 4733--4751}

\maketitle

\section{Introduction}\label{sec:introduction}

In the early 1970's, one-dimensional quantum integrable systems with
inverse-square long-range interactions appeared as a new class of
nontrivial solvable models, which is now generally called
Calogero--Moser--Sutherland (CMS)
models \cite{Calogero_1,Moser_1,Sutherland_1,Sutherland_2,vanDiejen_2}
in memory of the pioneers. 
Among the CMS models, the Calogero and the Sutherland 
models \cite{Calogero_1,Sutherland_1,Sutherland_2} are considered to be the
most typical models. The models describe many body systems confined by
the external harmonic well or the periodic boundary condition, which are
typical of the models in condensed matter physics.
In particular, the Sutherland model attracted many researchers because
its orthogonal basis were known to be the Jack polynomial 
\cite{Jack_1,Macdonald_1,Stanley_1} among physicists already in early 90's.
The theory of the Jack polynomial enabled a calculation of the exact
correlation functions of the Sutherland model
\cite{Ha_1,Kato_1,Uglov_1} and a related model in condensed matter
physics \cite{Kato_2}. 

The quantum integrability of the two models in a sense that they have
enough number of commutative conserved operators are explicitly shown
by the Dunkl--Cherednik operator formulations of a common 
structure to the two models~\cite{Dunkl_1,Cherednik_1,Polychronakos_1}.
The formulations are extended and generalized from the point of view of
the affine root systems
so as to cover a wide class
of the CMS models and to clarify relationships with other integrable
systems \cite{Komori_2,Komori_3}.
The celebrated Jack symmetric
polynomials~\cite{Jack_1,Stanley_1,Macdonald_1}
are the simultaneous eigenfunctions of conserved operators made of the
Cherednik operators of the Sutherland model. However, only a little had been
known about the symmetric simultaneous eigenfunctions of the conserved
operators of the Calogero model that are made of the Cherednik 
operators \cite{Ujino_2}.
Motivated by the Rodrigues formula for the Jack symmetric polynomial that was
found by Lapointe and Vinet~\cite{Lapointe_1,Lapointe_2},
we presented the Rodrigues formula for the Hi-Jack symmetric
(multivariable Hermite) polynomial~\cite{Lassalle_1} and 
identified it as the simultaneous eigenfunction of 
the conserved operators of the Calogero model%
~\cite{Ujino_1,Ujino_3,Ujino_4}. The multivariable Hermite polynomial is 
a one-parameter deformation of the Jack symmetric polynomial. They share
many common properties, which reflect the same algebraic structure of
the corresponding Dunkl--Cherednik operators. Moreover, the multivariable
Laguerre polynomials as well as the above multivariable Hermite polynomials
are investigated~\cite{Baker_0,Kakei_1,vanDiejen_1}.

To study the Calogero and Sutherland models including spin variables,
we need the non-symmetric simultaneous eigenvectors of the Cherednik
operators as the orthogonal basis of the orbital part of 
the eigenstate~\cite{Kato_1,Takemura_1,Takemura_2}.
Such a non-symmetric
simultaneous eigenfunction of the conserved operators of the Sutherland
model is known to be the non-symmetric Jack polynomial whose properties
are extensively studied in mathematical 
context~\cite{Opdam_1,Opdam_2,Knop_1,Sahi_1,Takamura_1}. On the other hand,
the simultaneous eigenfunction of the Calogero model
is identified as the non-symmetric multivariable Hermite
polynomial that is a one-parameter deformation of the non-symmetric Jack
polynomial~\cite{Baker_1}.
Some of the results for the non-symmetric Jack polynomials were
translated to the theory of the non-symmetric multivariable
Hermite and Laguerre polynomials~\cite{Baker_2,Nishino_1,Ujino_5,Ujino_6}.
As is similar to the symmetric polynomial case, however, 
less properties are clarified on
the non-symmetric multivariable Hermite and Laguerre polynomials 
than those of the non-symmetric Jack polynomials.

Recently, we investigated non-symmetric Jack and Macdonald
polynomials and their symmetrization and anti-symmetrization by
an algebraic formulation employing the Dunkl--Cherednik operators
together with the theory of root systems \cite{Nishino_4,Nishino_5}.
Through the method, algebraic constructions and
evaluations of square norms for non-symmetric, symmetric and anti-symmetric
multivariable polynomials can be treated in a unified way.
In this paper, we shall extend and apply the above method to the
multivariable Hermite and Laguerre polynomials. We shall present
algebraic constructions of the non-symmetric polynomials, 
symmetrizations and anti-symmetrizations, and evaluation of the square norms
of the Hermite and Laguerre cases,
which were not clarified in our previous 
works~\cite{Nishino_1,Ujino_1,Ujino_3,Ujino_4,Ujino_5,Ujino_6}.

The outline of the paper is as follows. In \sref{sec:Dunkl_operators},
we give a brief summary on the Dunkl--Cherednik operator formulation
for the Calogero models. In \sref{sec:polynomials},
the non-symmetric multivariable Hermite and Laguerre polynomials are 
introduced as the joint eigenvectors of the Cherednik operators.
We also introduce a notation based on the $A_{N-1}$-root system associated
with the finite-dimensional simple Lie algebra. 
In \sref{sec:Rodrigues}, the algebraic construction of
the non-symmetric multivariable Hermite and Laguerre polynomials are presented.
Square norms of the polynomials are calculated in an algebraic manner.
In \sref{sec:(anti-)symmetrization}, we construct symmetric and anti-symmetric
polynomials and compute their square norms.
The final section is devoted to a summary. A proof of a lemma
is presented in the appendix.

\section{Dunkl--Cherednik operators and the Calogero models}
\label{sec:Dunkl_operators}
We give a brief summary on the Dunkl--Cherednik operator formulation
of the $A_{N-1}$- and $B_{N}$-Calogero models~\cite{Olshanetsky_1},
which were named after the fact that the corresponding root systems appear
in their interaction terms. Actually, 
the $B_N$-Calogero model is the $C_N$-Calogero model at the same time, 
and reduces to the $D_N$-Calogero model by fixing the parameter $b=0$.
Thus the two models we shall study cover all the
Calogero models associated with root systems of classical simple
Lie algebras in the above mentioned sense.
The Hamiltonians of the Calogero models with distinguishable
particles~\cite{Polychronakos_1,Yamamoto_1,Baker_1} are expressed as
\begin{subequations}
  \begin{eqnarray}
    \fl \hat{\mathcal H}^{(A)} & = \frac{1}{2}\sum_{j=1}^{N}
      \bigl(-\frac{\partial^2}{\partial x_j^2}+\omega^{2}x_{j}^{2}\bigr)
      +\frac{1}{2}\sum_{\stackrel{\scriptstyle j,k=1}{j\neq k}}^{N}
      \frac{a^{2}-aK_{jk}}{(x_{j}-x_{k})^{2}},
    \label{eq:Calogero_Hamiltonian_A}\\
    \fl \hat{\mathcal H}^{(B)} & = \frac{1}{2}\sum_{j=1}^{N}
      \bigl(-\frac{\partial^2}{\partial x_j^2}
      +\omega^{2}x_{j}^{2}+\frac{b^2-b t_j}{x_j^2}\bigr) 
      {}+\frac{1}{2}\sum_{\stackrel{\scriptstyle j,k=1}{j\neq k}}^{N}
      \Bigl(\frac{a^{2}-aK_{jk}}{(x_{j}-x_{k})^{2}}
      +\frac{a^{2}-at_jt_kK_{jk}}{(x_{j}+x_{k})^{2}}\Bigr),
    \label{eq:Calogero_Hamiltonian_B}
  \end{eqnarray}
  \label{eq:Calogero_Hamiltonians}
\end{subequations}
where the coordinate exchange operator $K_{jk}$ and the reflection 
operator $t_j$ are defined as
\begin{eqnarray*}
  (K_{jk}f)(\cdots,x_{j},\cdots,x_{k},\cdots) = 
    f(\cdots,x_{k},\cdots,x_{j},\cdots),
  \\
  (t_j f)(\cdots,x_{j},\cdots) = f(\cdots,-x_{j},\cdots),\quad 
  j,k\in\{1,2,\cdots,N\},
\end{eqnarray*}
and we assume that the coupling parameters $a,b\in {\mathbb R_{\geq 0}}$.
In general, the eigenstates of the above Calogero Hamiltonians 
\eref{eq:Calogero_Hamiltonians} are non-symmetric 
with respect to exchanges of particle indices.
That is why we have called them the models with distinguishable particles.
The eigenfunctions of the Calogero models are expressed as
the products of inhomogeneous non-symmetric multivariable polynomials, 
namely the non-symmetric multivariable Hermite and Laguerre polynomials%
~\cite{Baker_1,Dunkl_2,Ujino_6,Nishino_1}, and the reference states.
To study the polynomial part of such eigenfunctions, we introduce
transformed Hamiltonians whose eigenvectors are polynomials,
\begin{equation}
  {\mathcal H}^{(A,B)} \Define \bigl(\phi_{\rm g}^{(A,B)}(x)
  \bigr)^{-1}\Mmae\circ
  \bigl(\hat{\mathcal{H}}^{(A,B)}-E_{\rm g}^{(A,B)}\bigr)
  \circ\phi_{\rm g}^{(A,B)}(x),
  \label{eq:transformed_Hamiltonian}
\end{equation}
where
\begin{subequations}
  \begin{eqnarray}
    \eqalign{
      \phi_{\rm g}^{(A)}(x) = \Mmae
      \prod_{1\leq j<k\leq N}\Mmae |x_{j}-x_{k}|^{a}
      \exp\Bigl(-\frac{1}{2}\omega\sum_{m=1}^{N}x_{m}^{2}\Bigr),\\
      E_{\rm g}^{(A)} = \frac{1}{2}\omega N\bigl(Na+(1-a)\bigr),
    }\label{eq:Calogero_ground_state_A}\\
    \eqalign{
      \phi_{\rm g}^{(B)}(x)= \Mmae \prod_{1\leq j<k\leq N}\Mmae
      |x_{j}^2-x_{k}^2|^{a} \prod_{l=1}^{N}|x_l|^b
      \exp\Bigl(-\frac{1}{2}\omega\sum_{m=1}^{N}x_{m}^{2}\Bigr),\\
      E_{\rm g}^{(B)} = \frac{1}{2}\omega N\bigl(2Na+(1-2a)+2b\bigr).
    }
    \label{eq:Calogero_ground_state_B}
  \end{eqnarray}
  \label{eq:Calogero_ground_states}
\end{subequations}
The above reference states and their eigenvalues are known to be the ground
states and the ground state energies for the $A_{N-1}$- and 
the $B_N$-Calogero models with distinguishable particles and the bosonic
particles.
In the following, we call \eref{eq:transformed_Hamiltonian}
instead of \eref{eq:Calogero_Hamiltonians} the Calogero Hamiltonians.

Let ${\mathbb C}[x]$ be the polynomial ring with $N$ variables over 
${\mathbb C}$. We deal with the eigenfunctions for the original Calogero
Hamiltonians $\hat{\mathcal H}^{(A,B)}$ in the spaces ${\mathbb C}[x]
\phi_{\rm g}^{(A,B)}=\{f(x)\phi_{\rm g}^{(A,B)}(x)|f\in\mathbb{C}[x]\}$ 
with the following canonical inner products,
\[
  (\varphi,\psi)\Define\int^{\infty}_{-\infty}\prod_{j=1}^{N}
  \d x_j \overline{\varphi(x)}\psi(x),
  \quad \mbox{for }\varphi, \psi \in {\mathbb C}[x]
  \phi_{\rm g}^{(A,B)},
\]
where $\overline{\varphi(x)}$
means the complex conjugate of $\varphi(x)$.
On the other hand, the transformed 
Hamiltonians \eref{eq:transformed_Hamiltonian} are hermitian with respect
to the inner product on ${\mathbb C}[x]$,
\begin{equation}
  \langle f, g\rangle_{(A,B)}
    \Define\int_{-\infty}^{\infty}\prod_{j=1}^{N}\d x_{j}
    |\phi_{\rm g}^{(A,B)}(x)|^{2}\overline{f(x)}g(x), 
    \quad \text{ for }\, f,g \in {\mathbb C}[x],
  \label{eq:inner_product}
\end{equation}
which are induced from $(\cdot,\cdot)$ and the transformation %
\eref{eq:transformed_Hamiltonian}.
Thus the reference states \eref{eq:Calogero_ground_states} correspond to
the weight functions in the above inner products %
$\langle\cdot,\cdot\rangle_{(A,B)}$.
The commuting conserved operators for the Calogero Hamiltonians
are known to be the Cherednik operators. To show this,
we need to introduce the Dunkl operators $\nabla_{j}^{(A,B)}\in\End(
\mathbb{C}[x])$~\cite{Dunkl_1},
\begin{eqnarray*}
  \fl\nabla_{j}^{(A)} \Define \frac{\partial}{\partial x_{j}}
  + a\sum_{\stackrel{\scriptstyle k=1}{k\neq j}}^{N}\frac{1}{x_{j}-x_{k}}
  (1-K_{jk}),
  \\
  \fl\nabla_{j}^{(B)} \Define \frac{\partial}{\partial x_{j}}
  + a\sum_{\stackrel{\scriptstyle k=1}{k\neq j}}^{N}\Bigl(
  \frac{1}{x_{j}-x_{k}}(1-K_{jk})
  + \frac{1}{x_{j}+x_{k}}(1-t_jt_kK_{jk})\Bigr) + \frac{b}{x_j}(1-t_j),
\end{eqnarray*}
and the creation-like and annihilation-like operators 
$\alpha_{l}^{(A,B)\dagger},\alpha_{l}^{(A,B)}\in\End(\mathbb{C}[x])$
for the Calogero models,
\[
  \alpha_{l}^{(A,B)\dagger}\Define
  x_{l}-\frac{1}{2\omega}\nabla_{l}^{(A,B)},\quad
  \alpha_{l}^{(A,B)}=\frac{1}{2\omega}\nabla_{l}^{(A,B)}, 
\]
where the superscript $\dagger$ on any operator denotes 
its hermitian conjugate with respect
to the inner product \eref{eq:inner_product}.
From these operators, two sets of hermitian and commutative differential
operators $d_{j}^{(A,B)}\in\End(\mathbb{C}[x])$, 
$[d_j^{(A,B)},d_k^{(A,B)}]=0$, \cite{Kakei_1,Baker_2} are constructed by
\begin{eqnarray*}
  d_{j}^{(A)} \Define 2\omega\alpha_{j}^{(A)\dagger}\alpha_{j}^{(A)}
    +a\!\!\!\sum_{k=j+1}^{N}K_{jk},\\
  d_{j}^{(B)} \Define 2\omega\alpha_{j}^{(B)\dagger}\alpha_{j}^{(B)}
    +a\!\!\!\sum_{k=j+1}^{N}(1+t_jt_k)K_{jk}+bt_j.
\end{eqnarray*}
We call them the Cherednik 
operators~\cite{Cherednik_1,Cherednik_2,Cherednik_3,Kakei_1}.
The Cherednik operators and the exchange and reflection operators satisfy
\begin{subequations}
  \begin{eqnarray}
    \eqalign{
      d_{l}^{(A)}K_{l}-K_{l}d_{l+1}^{(A)}=a, \quad
      d_{l+1}^{(A)}K_{l}-K_{l}d_{l}^{(A)}=-a, \\
      [d_{l}^{(A)},K_{m}]=0,\quad 
      \text{ for }\, l\neq m,m+1,
    }
    \label{eq:commutation_Cherednik_exchange_reflection_A}\\
    \eqalign{
      d_{l}^{(B)}K_{l}-K_{l}d_{l+1}^{(B)}=a(1+t_lt_{l+1}),\\
      d_{l+1}^{(B)}K_{l}-K_{l}d_{l}^{(B)}=-a(1+t_lt_{l+1}),\\
      [d_{l}^{(B)},K_{m}]=0,
      \quad\text{ for }\, l\neq m,m+1,\\
      [d_{l}^{(B)},t_{m}]=0,
    }
    \label{eq:commutation_Cherednik_exchange_reflection_B}
  \end{eqnarray}
  \label{eq:commutation_Cherednik_exchange_reflection}
\end{subequations}
where the exchange operators $K_{l,l+1}$ for
$l\in\{1,2,\cdots,N-1\}$ are denoted by $K_l$.
In terms of the Cherednik operators, 
the Calogero Hamiltonians \eref{eq:transformed_Hamiltonian}
can be expressed as
\begin{subequations}
  \begin{eqnarray}
    {\mathcal H}^{(A)}
      = \omega\sum_{l=1}^{N}\Bigl(d_{l}^{(A)}-\frac{1}{2}a(N-1)\Bigr),
    \label{eq:Cherednik2Hamiltonian_A}\\
    {\mathcal H}^{(B)}
      = \omega\sum_{l=1}^{N}\Bigl(d_{l}^{(B)}-a(N-1)-b\Bigr).
    \label{eq:Cherednik2Hamiltonian_B}
  \end{eqnarray}
  \label{eq:Cherednik2Hamiltonian}
\end{subequations}
Thus we conclude that
the Cherednik operators $\{d_{l}^{(A,B)}|l=1,2,\cdots,N\}$
give the sets of commutative conserved operators of the Calogero models.
The last formula among 
equations \eref{eq:commutation_Cherednik_exchange_reflection_B}
and the $B_{N}$-Calogero Hamiltonian \eref{eq:Cherednik2Hamiltonian_B}
imply that the parity of each variable is a good quantum number of
the $B_N$-Calogero model with distinguishable particles.

\section{Non-symmetric multivariable Hermite and Laguerre polynomials}
\label{sec:polynomials}

The Cherednik operators define inhomogeneous multivariable polynomials
as their joint polynomial eigenfunctions, which are nothing but 
the non-symmetric multivariable Hermite and Laguerre polynomials 
that form orthogonal bases of the polynomial ring ${\mathbb C}[x]$%
~\cite{Baker_1,Dunkl_2,Ujino_6,Nishino_1}.

In order to investigate such polynomial eigenfunctions, we need mathematical
preparations for a root system and the associated Weyl 
group~\cite{Humphreys}.
Let $\Check{I}=\{1,2,\cdots,N-1\}$ and $I=\{1,2,\cdots,N\}$
be sets of indices and
let $V$ be an $N$-dimensional real vector space with positive
definite bilinear form $\langle\cdot,\cdot\rangle$.
We take an orthogonal basis $\{\varepsilon_{j}|j\in I\}$ of $V$
such that $\langle\varepsilon_{j},\varepsilon_{k}\rangle=\delta_{jk}$.
We realize the $A_{N-1}$-type root system $R$ 
associated with the simple Lie algebra of type $A_{N-1}$ as
\[
  R=\{\varepsilon_{j}-\varepsilon_{k}|j,k\in I,j\neq k\}(\subset V).
\]
A root basis of $R$ is defined by
\[
  \Pi\Define\{\alpha_{j}=\varepsilon_{j}-\varepsilon_{j+1}|
  j\in \check{I}\},
\]
whose elements are called simple roots.
We denote by $R_{+}$ the set of positive roots relative to $\Pi$ and
$R_{-}=-R_{+}$. The root lattice $Q$ is defined by
$Q\Define\bigoplus_{j\in\check{I}}{\mathbb Z}\alpha_j$ and the positive
root lattice $Q_{+}$ is defined by replacing ${\mathbb Z}$ with
${\mathbb Z}_{\geq 0}$.

We consider a reflection on $V$ with respect to 
the hyperplane that is orthogonal to a root $\alpha\in R$,
and indicate it by
$s_{\alpha}(\mu)\Define\mu-\langle\alpha^{\vee},\mu\rangle\alpha$,
where $\alpha^{\vee}\Define 2\alpha/\langle\alpha,\alpha\rangle$ 
is a coroot corresponding to $\alpha\in R$.
The reflections $\{s_{j}\Define s_{\alpha_{j}}|\alpha_{j}\in \Pi\}$ 
generate the $A_{N-1}$-type Weyl group $W$
which is isomorphic to the symmetric group ${\mathfrak S}_{N}$,
$W\simeq {\mathfrak S}_{N}$.
For each $w\in W$, we define the following set of positive roots:
$R_{w}\Define R_{+}\cap w^{-1}R_{-}$. We denote by $\ell(w)$
the length of $w\in W$ defined by $\ell(w)\Define|R_w|$.
When $w\in W$ is written as a product of simple reflections, e.g.,
$w=s_{j_{k}}\cdots s_{j_{2}}s_{j_{1}}$, the length $\ell(w)$ gives the
smallest $k$ in such expressions. We call an expression
$w=s_{j_{l}}\cdots s_{j_{2}}s_{j_{1}}$, $l=\ell(w)$, reduced.
If we take the above reduced expression, 
the set $R_{w}$ is expressed by
\[
  R_{w}=\{\alpha_{j_{1}},s_{j_{1}}(\alpha_{j_{2}}),\cdots,
          s_{j_{1}}s_{j_{2}}\cdots s_{j_{l-1}}(\alpha_{j_{l}})\}.
\]
Though reduced expressions may not be unique for each $w\in W$, it 
is known that the above set $R_{w}$ is unique as a set for each
$w\in W$ \cite{Humphreys}.
 
We introduce lattices 
$P\Define \bigoplus_{j\in I}{\mathbb Z}_{\geq 0}\varepsilon_{j}$
and $P_{+}\Define 
\{\mu=\sum_{j\in I}\mu_{j}\varepsilon_{j}\in P
  |\mu_{1}\geq\mu_{2}\geq\cdots\geq\mu_{N}\geq 0\}$
whose elements are called a composition and a partition, respectively.
The lattice $P$ is $W$-stable.
The degree of the composition and partition is denoted by $|\mu|\Define
\sum_{j\in I}\mu_j $.
Let $W(\mu)\Define\{w(\mu)|w\in W\}$ be the $W$-orbit of $\mu\in P$.
In a $W$-orbit $W(\mu)$, 
there exists a unique partition $\mu^{+}\in P_{+}$
such that $\mu=w(\mu^{+})\in P\, (w\in W)$.
We define
\begin{eqnarray*}
  \rho\Define 
  \frac{1}{2}\sum_{\alpha\in R_{+}}\alpha
  =\frac{1}{2}\sum_{j\in I}(N-2j+1)\varepsilon_{j},\quad
  1^N\Define\sum_{j\in I}\varepsilon_j,\\
  \delta\Define\sum_{j\in I}(N-j)\varepsilon_{j}
  =\rho+\frac{1}{2}(N-1)1^N .
\end{eqnarray*}
In order to deal with the eigenvalues of the Cherednik operators in
terms of the lattice $P$, we introduce the following operators,
\[
  d^{(A,B)\lambda}\Define\sum_{j\in I}\lambda_j d^{(A,B)}_{j}, \quad
  t^{\lambda} \Define t_1^{\lambda_1}t_2^{\lambda_2}\cdots t_N^{\lambda_N},
  \quad \lambda\in P,
\]
which relate the Cherednik and reflection operators with the lattice $P$.

We identify the elements of the lattice $P$ with 
those of the polynomial ring with $N$ variables
over ${\mathbb C}$, 
$x^{\mu}\Define 
 x_{1}^{\mu_{1}}x_{2}^{\mu_{2}}\cdots x_{N}^{\mu_{N}}
 \in{\mathbb C}[x]$.
Then the action of the coordinate exchange operators 
$\{K_{j}|j\in \Check{I}\}$ on ${\mathbb C}[x]$ are written as
\[
  K_{j}(x^{\mu})=x^{s_{j}(\mu)},\quad
  \text{ for } x^{\mu}\in {\mathbb C}[x].
\]
We denote the ($W$-)symmetric and ($W$-)anti-symmetric polynomial rings
over $\mathbb{C}$ by $\mathbb{C}[x]^{\pm W}$.
On the other hand,
the action of the reflection operators on ${\mathbb C}[x]$ is expressed as
\[
  t_j(x^{\mu})=(-1)^{\langle\varepsilon_j,\mu\rangle}x^{\mu},\quad
  t^{\alpha_{j}^{\vee}}(x^{\mu})
  =(-1)^{\langle\alpha_j^{\vee},\mu\rangle}x^{\mu},\quad
  \text{ for } x^{\mu}\in {\mathbb C}[x].
\]
We shall use such notations quite often in the following.

We denote the shortest element of $W$
such that $w_{\mu}^{-1}(\mu)\in P_{+}$ by $w_{\mu}$ and define 
$\rho(\mu)\Define w_{\mu}(\rho)$
and $\delta(\mu)\Define w_{\mu}(\delta)$. The definitions of
the (monic) non-symmetric multivariable Hermite and Laguerre polynomials,
$h_{\mu}^{(A,B)} \in {\mathbb C}[x]$, $\mu\in P$, 
as the joint eigenvectors for the commutative Cherednik 
operators $\{d^{(A,B)\lambda}\}$ are given by
\begin{subequations}
  \begin{eqnarray}
    \eqalign{
      h_{\mu}^{(A)}(x) & = x^{\mu}
        + \Mmae \sum_{\nu \preceq \mu\atop {\rm or} \; |\nu|<|\mu|}\Mmae
        v_{\mu\nu}^{(A)}(a,\frac{1}{2\omega})x^{\nu}, \\
      d^{(A)\lambda}h_{\mu}^{(A)} & = 
      \langle\lambda,\mu+a\rho(\mu)+\frac{1}{2}a(N-1)1^N\rangle 
      h_{\mu}^{(A)} \\
      & =\langle\lambda,\mu+a\delta(\mu)\rangle h_{\mu}^{(A)},
    }
    \label{eq:nonsym-Hermite}\\
    \eqalign{
      h_{\mu}^{(B)}(x) & = x^{\mu}
        +\Mmae\sum_{\nu \preceq \mu\atop {\rm or} \; |\nu|<|\mu|}\Mmae
        v_{\mu\nu}^{(B)}(a,b,\frac{1}{2\omega})x^{\nu}, \\
      d^{(B)\lambda}h_{\mu}^{(B)} & =
      \langle\lambda,\mu+ 2a\rho(\mu)+(a(N-1)+b)1^N\rangle
        h_{\mu}^{(B)} \\
      & = \langle\lambda,\mu+\rho_k^{(B)}(\mu)\rangle h_{\mu}^{(B)},
    }
    \label{eq:non-symmetric_Laguerre}
  \end{eqnarray}
  \label{eq:non-symmetric_polynomials}
\end{subequations}
where
\[
  \rho_k^{(B)}\Define\sum_{j\in I}\bigl(2a(N-j)+b\bigr)\varepsilon_j,
  \quad \rho^{(B)}_k(\mu)\Define w_{\mu}(\rho^{(B)}_k).
\]
The triangularity is defined by the order $\preceq$ on $P$:
\begin{equation}
  \nu\preceq \mu \quad (\nu, \mu\in P) \Leftrightarrow 
  \cases{
    \nu^{+}\dominance \mu^{+} & $\nu\not\in W(\mu^{+})$, \\
    \mu - \nu \in Q_+ & $\nu\in W(\mu^{+})$.
  }\label{eq:bruhat}
\end{equation}
Here, the symbol $\dominance$ denotes
the dominance order among partitions
\[
  \nu\dominance\mu \quad (\mu, \nu \in P_+)
  \Leftrightarrow \mu\neq\lambda, \; |\mu|=|\nu|
  \text{ and }\sum_{k=1}^{l}\nu_{k}\leq\sum_{k=1}^{l}\mu_{k},
\]
for all $l\in I$. We should note that the non-symmetric multivariable
Laguerre polynomial is the joint eigenvector of the reflection
operators $t_j$, $j\in I$, 
\[
  t^{\lambda}h_{\mu}^{(B)} 
  = (-1)^{\langle\lambda,\mu\rangle}h_{\mu}^{(B)},
\]
and the parity with respect to each variable is 
a quantum number of the $B_N$-Calogero
models with distinguishable particles. The above formula 
tells that the parity of $h^{(B)}_\mu$
with respect to a variable $x_j$ is $(-1)^{\mu_j}$.

Since $d^{(A,B)\lambda}$ are hermitian operators with respect to the
inner products \eref{eq:inner_product},
\[
  \langle f,d^{(A,B)\lambda}g \rangle_{(A,B)} 
  = \langle d^{(A,B)\lambda}f,g \rangle_{(A,B)},
\]
and all the simultaneous eigenspaces of the Cherednik operators 
$\{d^{(A,B)\lambda}\}$ are one-dimensional in the sense
that the eigenvalues of $\{d^{(A,B)\lambda}\}$ are non-degenerate,
which proves that the polynomials $h_{\mu}^{(A,B)}$ are orthogonal 
with respect to the inner product,
i.e., $\langle h_{\mu}^{(A,B)}, h_{\nu}^{(A,B)} \rangle_{(A,B)}
 =\delta_{\mu,\nu}\|h_{\mu}^{(A,B)}\|^{2}$.
In fact, the non-symmetric multivariable Hermite and Laguerre polynomials
form orthogonal bases in ${\mathbb C}[x]$.
We readily confirm that the polynomials \eref{eq:non-symmetric_polynomials}
are generally non-symmetric under exchanges of variables $\{x_{j}\}$.

We should note a connection of 
the action of the Weyl group and the above definition of 
the order $\preceq$ \eref{eq:bruhat} for
compositions in the same $W$-orbit.
Let us compare $s_j(\mu)$ and $\mu$ by the order $\preceq$. 
From the definition of the reflection, we have
\[
  s_j(\mu)-\mu = -\langle\alpha_j^{\vee},\mu\rangle\alpha_j .
\]
Thus we conclude $\mu\succeq s_j(\mu)$ if
$\langle\alpha_j^{\vee},\mu\rangle \geq 0$.
When one of the reduced expressions of $w_{\mu}$ is given by 
$s_{j_l}\cdots s_{j_2}s_{j_1}$, $(l=\ell(w_{\mu}))$,
we can confirm the following relation,
\begin{equation}
  \eqalign{
    \mu=\mu^{(l)}\prec\mu^{(l-1)}\prec\cdots\prec\mu^{(0)}=\mu^+,\\
    \mu^{(n)} \Define s_{j_n}\cdots s_{j_2}s_{j_1}(\mu^+),
    \quad n\in\{0,1,2,\cdots,l\},
  }
  \label{eq:ordered_sequence}
\end{equation}
using the fact $\langle \alpha^{\vee},\mu^+\rangle\geq 0$, ${}^\forall
\alpha\in R_{w_\mu}\subseteq R_+$, $\mu^+\in P_+$. 
We shall use this relation in the algebraic construction of the polynomials
in the next section.

\section{Rodrigues formula}\label{sec:Rodrigues}

We shall present the Rodrigues formulas for the non-symmetric multivariable
Hermite and Laguerre polynomials $h_{\mu}^{(A,B)}$.
In order to calculate the
square norms of the polynomials, they should be monic in the sense that
the coefficients of the top terms are unity.
However, polynomials generated by the Rodrigues formulas
presented in our previous works~\cite{Nishino_1,Ujino_5} were not monic.
Here we show the Rodrigues formulas that generate the monic polynomials.

We introduce the Knop--Sahi operators $\{e^{(A,B)},e^{(A,B)\dagger}\}$ 
\cite{Knop_1} and 
the braid operators $\{S_{j}^{(A,B)}|j\in \Check{I}\}$ defined by
\begin{eqnarray}
  \eqalign{
    e^{(A,B)}\Define \alpha_{1}^{(A,B)}K_{1}K_{2}\cdots K_{N-1},\\
    e^{(A,B)\dagger} =
    K_{N-1}\cdots K_{2}K_{1}\alpha_{1}^{(A,B)\dagger},
  }\label{eq:Knop-Sahi_ops} \\
  S_{j}^{(A,B)}\Define [K_{j},d_{j}^{(A,B)}].
  \label{eq:braid_op}
\end{eqnarray}
The operators $\{e^{(A)},e^{(A)\dagger}\}$ were first introduced 
by Baker and Forrester~\cite{Baker_1}.
The Knop--Sahi operators and the braid operators satisfy
the following relations:
\begin{equation}
  \eqalign{
    S_{j}^{(A,B)}S_{j+1}^{(A,B)}S_{j}^{(A,B)}
    =S_{j+1}^{(A,B)}S_{j}^{(A,B)}S_{j+1}^{(A,B)}, \quad
    \text{ for }\, 1\leq j\leq N-2,\\
    S_{j}^{(A,B)}S_{k}^{(A,B)}=S_{k}^{(A,B)}S_{j}^{(A,B)}, \quad
    \mbox{ for }\, |j-k|\geq 2,\\
    t_jS_{j}^{(B)}t_{j+1}S_{j}^{(B)} = S_{j}^{(B)}t_{j+1}S_{j}^{(B)}t_j,
    \\
    S_{j}^{(A,B)}e^{(A,B)\dagger}=e^{(A,B)\dagger}S_{j+1}^{(A,B)},\quad
    \text{ for }\, 1\leq j\leq N-2,\\
    S_{N-1}^{(A,B)}(e^{(A,B)\dagger})^{2}
    =(e^{(A,B)\dagger})^{2}S_{1}^{(A,B)}, \\
    (S_{j}^{(A)})^{2}=a^{2}-(d_{j}^{(A)}-d_{j+1}^{(A)})^{2}, \\
    (S_{j}^{(B)})^{2}=2a^{2}(1+t_jt_{j+1})
    -(d_{j}^{(B)}-d_{j+1}^{(B)})^{2}, \\
    S_{j}^{(A,B)\dagger}=-S_{j}^{(A,B)},
    e^{(A,B)\dagger}e^{(A,B)}=\frac{1}{2\omega}d_{N}^{(A,B)},
  }
  \label{eq:S-e-prop}
\end{equation}
and
\begin{equation}
  \eqalign{
    S_{j}^{(A,B)}d^{(A,B)\lambda}=d^{(A,B)s_j(\lambda)}S_{j}^{(A,B)}, \quad
    S_{j}^{(B)}t^{\lambda}=t^{s_j(\lambda)}S_{j}^{(B)},\\ 
    d^{(A,B)\lambda}e^{(A,B)\dagger}=e^{(A,B)\dagger}\bigl(
    d^{(A,B)s_1s_2\cdots s_{N-1}(\lambda)}+\langle\lambda,\varepsilon_N\rangle
    \bigr).
  }
  \label{eq:S-e-d-prop}
\end{equation}
The first relation in \eref{eq:S-e-prop} is called braid relation.
The relations \eref{eq:S-e-d-prop} 
indicates that the operators $\{S_{j}^{(A,B)\dagger},e^{(A,B)\dagger}\}$
intertwine the simultaneous eigenspaces of $\{d^{(A,B)\lambda}\}$.
We define
the raising operators $\{A_{\mu}^{(A,B)\dagger}|\mu\in P_{+}\}$ by
\begin{equation}
  \eqalign{
    A_{\mu}^{(A,B)\dagger}\Define 
    (A_{1}^{(A,B)\dagger})^{\mu_{1}-\mu_{2}}
    (A_{2}^{(A,B)\dagger})^{\mu_{2}-\mu_{3}}\cdots
    (A_{N}^{(A,B)\dagger})^{\mu_{N}}, \nn\\
    A_{j}^{(A,B)\dagger}
    \Define (S_{j}^{(A,B)}S_{j+1}^{(A,B)}\cdots 
    S_{N-1}^{(A,B)}e^{(A,B)\dagger})^{j},\quad
    \text{ for }\, j\in I.
  }\label{eq:raising_operators}
\end{equation}
Equations \eref{eq:S-e-prop} and \eref{eq:S-e-d-prop} yield the following
relations of the raising operators,
\begin{equation}
  \eqalign{
    d^{(A,B)\lambda}A_{\mu}^{(A,B)\dagger}=A_{\mu}^{(A,B)\dagger}
    (d^{(A,B)\lambda}+\langle\lambda,\mu\rangle),\\
    t^{\lambda}A_{\mu}^{(B)\dagger}=(-1)^{\langle\lambda,\mu\rangle}
    A_{\mu}^{(B)\dagger}t^{\lambda},\\
    \big[A_{\mu}^{(A,B)\dagger},A_{\nu}^{(A,B)\dagger}\big]=0,\quad
    \text{ for }\, \mu,\nu\in P_+,
  }\label{eq:raising_commutators}
\end{equation}
which lead to the Rodrigues formula for 
the non-symmetric multivariable Hermite and Laguerre
polynomials that are identified with partitions, 
$h_{\mu}^{(A,B)}$, $\mu\in P_{+}$.
\begin{proposition}[cf.~\cite{Nishino_1,Ujino_6}]
The non-symmetric multivariable Hermite and Laguerre 
polynomials $h_{\mu}^{(A,B)}$
with a partition $\mu\in P_{+}$ are algebraically
constructed by applying the raising operators $A_{\mu}^{(A,B)\dagger}$
to $h_0^{(A,B)}=1$,
\begin{equation}
  h_{\mu}^{(A,B)}
  ={c_{\mu}^{(A,B)}}^{-1}A_{\mu}^{(A,B)\dagger}h_{0}^{(A,B)}, \\
  \label{eq:Rodrigues_partition}
\end{equation}
where the coefficients of the top term $c_{\mu}^{(A,B)}$ are 
given by
\begin{subequations}
  \begin{eqnarray}
    c_{\mu}^{(A)} \Define \prod_{\alpha\in R_{+}}\prod_{l=1}%
    ^{\langle\alpha^{\vee},\mu\rangle}\bigl(
    l+a\langle\alpha^{\vee},\rho\rangle\bigr), 
    \label{eq:dominant_coeff_A}\\
    c_{\mu}^{(B)} \Define \prod_{\alpha\in R_{+}}\prod_{l=1}%
    ^{\langle\alpha^{\vee},\mu\rangle}\bigl(
    l+2a\langle\alpha^{\vee},\rho\rangle\bigr) .
    \label{eq:dominant_coeff_B}
  \end{eqnarray}
  \label{eq:dominant_coeffs}
\end{subequations}
\label{pr:Rodrigues_partition}
\end{proposition}

To calculate the coefficients, we need to
know the action of the coordinate exchange operators on the polynomials.
\begin{lemma}[cf.~\cite{Baker_2,Knop_1}]
Applying the coordinate exchange operators
$\{K_{j}|j\in \check{I}\}$ to the non-symmetric Hermite and Laguerre
polynomials $h_{\mu}^{(A,B)}\in{\mathbb C}[x], (\mu\in P)$, 
we find that
\begin{eqnarray*}
    \fl K_{j}h_{\mu}^{(A)}
    = \cases{
      \dfrac{a}{\langle\alpha_{j}^{\vee},\mu+ a\rho(\mu)\rangle}h_{\mu}^{(A)}
        +h_{s_{j}(\mu)}^{(A)}, & if $\langle\alpha_{j}^{\vee},\mu\rangle<0$,\\
      h_{\mu}^{(A)}, & if $\langle\alpha_{j}^{\vee},\mu\rangle=0$,\\
      \dfrac{a}{\langle\alpha_{j}^{\vee},\mu+ a\rho(\mu)\rangle}h_{\mu}^{(A)}
      +\Bigl(1-\dfrac{a^{2}}{\langle\alpha_{j}^{\vee},
      \mu+ a\rho(\mu)\rangle^{2}}\Bigr)h_{s_{j}(\mu)}^{(A)}, &
      if $\langle\alpha_{j}^{\vee},\mu\rangle>0$,
    } \\
    \fl K_{j}h_{\mu}^{(B)}
    = \cases{
      \dfrac{a(1+(-1)^{\langle\alpha^{\vee}_{j},\mu\rangle})}%
        {\langle\alpha_{j}^{\vee},\mu + 2a\rho(\mu)\rangle}h_{\mu}^{(B)}
        +h_{s_{j}(\mu)}^{(B)}, & if $\langle\alpha_{j}^{\vee},\mu\rangle<0$,\\
      h_{\mu}^{(B)}, & if $\langle\alpha_{j}^{\vee},\mu\rangle=0$,\\
      \dfrac{a(1+(-1)^{\langle\alpha^{\vee}_{j},\mu\rangle})}
        {\langle\alpha_{j}^{\vee},\mu+ 2a\rho(\mu)\rangle}h_{\mu}^{(B)}
        +\Bigl(1-\dfrac{a^{2}(%
        1+(-1)^{\langle\alpha^{\vee}_{j},\mu\rangle})^{2}}
        {\langle\alpha_{j}^{\vee},\mu + 2a\rho(\mu)\rangle^{2}}\Bigr)
        h_{s_{j}(\mu)}^{(B)},& if $\langle\alpha_{j}^{\vee},\mu\rangle>0$.
    }
\end{eqnarray*}
\label{lm:K_expansions}
\end{lemma}
\Proof
The proof of the lemma is straightforward by using the definitions 
and orthogonality of the polynomials and the commutation relation
\eref{eq:commutation_Cherednik_exchange_reflection}. \QED

\noindent
{\bf Proof of Proposition \ref{pr:Rodrigues_partition}}\;
Let $\tilde{h}_{\mu}^{(A,B)}\Define
A_{\mu}^{(A,B)\dagger}h_{0}^{(A,B)}$.
By a straightforward calculation using \eref{eq:raising_commutators},
we can confirm
\begin{eqnarray*}
  d^{(A)\lambda}\tilde{h}_{\mu}^{(A)}
  & = A_{\mu}^{(A)\dagger}(d^{(A)\lambda}+\langle\lambda,
  \mu\rangle)h_{0}^{(A)} \\
  & = \langle\lambda,\mu+a\rho(\mu)+\frac{1}{2}a(N-1)1^N\rangle
  \tilde{h}_{\mu}^{(A)},\\
  d^{(B)\lambda}\tilde{h}_{\mu}^{(B)}
  & = A_{\mu}^{(B)\dagger}(d^{(B)\lambda}+\langle\lambda,
  \mu\rangle)h_{0}^{(B)} \\
  & = \langle\lambda,\mu+2a\rho(\mu)+(a(N-1)+b)1^N\rangle
  \tilde{h}_{\mu}^{(B)},
\end{eqnarray*}
which are nothing but the second and the forth relation 
in the definitions of the
non-symmetric multivariable Hermite and Laguerre polynomials
\eref{eq:non-symmetric_polynomials}.
Since all the simultaneous eigenspaces of the Cherednik 
operators $\{d^{(A,B)\lambda}\}$ are
one-dimensional, we can identify $\tilde{h}_{\mu}^{(A,B)}$ with the 
non-symmetric multivariable Hermite and Laguerre polynomials
$h_{\mu}^{(A,B)}$, $(\mu\in P_{+})$ up to constant multiplicative
coefficients $c_{\mu}^{(A,B)}$. 

{}From the lemma above 
and the definition of the braid operators \eref{eq:braid_op},
it is straightforward to compute the action of the braid operators
$S_j^{(A,B)}$ on $h_{\mu}^{(A,B)}$.
\begin{subequations}
  \begin{eqnarray}
    \fl \eqalign{
      S_{j}^{(A)}h_{\mu}^{(A)}
      = \cases{
        \dfrac{a}{\langle\alpha_{j}^{\vee},\mu
        + a\rho(\mu)\rangle}h_{s_j(\mu)}^{(A)}, & 
        if $\langle\alpha_{j}^{\vee},\mu\rangle<0$,\\
        0, & if $\langle\alpha_{j}^{\vee},\mu\rangle=0$,\\
        \dfrac{\langle\alpha_{j}^{\vee},
        \mu+ a\rho(\mu)\rangle^2-a^{2}}{\langle\alpha_{j}^{\vee},
        \mu+ a\rho(\mu)\rangle}h_{s_{j}(\mu)}^{(A)}, &
        if $\langle\alpha_{j}^{\vee},\mu\rangle>0$,
      }
    }
    \label{eq:S_expansion_A}
    \\
    \fl \eqalign{
      S_{j}^{(B)}h_{\mu}^{(B)} 
      = \cases{
        \langle\alpha_{j}^{\vee},\mu + 2a\rho(\mu)\rangle 
        h_{s_{j}(\mu)}^{(B)}, & if $\langle\alpha_{j}^{\vee},\mu\rangle<0$,\\
        0, & if $\langle\alpha_{j}^{\vee},\mu\rangle=0$,\\
          \dfrac{\langle\alpha_{j}^{\vee},\mu + 2a\rho(\mu)\rangle^{2}
          -a^{2}(1+(-1)^{\langle\alpha^{\vee}_{j},\mu\rangle})^{2}}
          {\langle\alpha_{j}^{\vee},\mu + 2a\rho(\mu)\rangle}
          h_{s_{j}(\mu)}^{(B)},& if $\langle\alpha_{j}^{\vee},\mu\rangle>0$.
      }
    }
    \label{eq:S_expansion_B}
  \end{eqnarray}
  \label{eq:S_expansions}
\end{subequations}
Then we can confirm that the coefficients 
$c_{\mu}^{(A,B)}$ are given by \eref{eq:dominant_coeffs} 
by calculations using \eref{eq:non-symmetric_polynomials} and
\eref{eq:S_expansions}. \QED

Since we have the braid operators $S_j^{(A,B)}$ that operate 
on the polynomials $h^{(A,B)}_{\mu}$ 
and generate polynomials $h^{(A,B)}_{s_j(\mu)}$, all we have to
do to construct the non-symmetric multivariable Hermite 
and Laguerre polynomials with a general composition $\mu\in P$ lying in 
$W(\mu^{+})$ is to apply the braid operators to 
the polynomials $h_{\mu^{+}}^{(A,B)}$, $\mu^+\in P_+$.
\begin{proposition}
Let $w_\mu =s_{j_l}\cdots s_{j_2}s_{j_1}$
be one of the reduced expressions of $w_{\mu}$
and let $S_{w_\mu}$ be defined by 
$S_{w_\mu}\Define S_{j_l}\cdots S_{j_2}S_{j_1}$.
Then the non-symmetric multivariable Hermite and Laguerre polynomials with 
a composition $\mu\in P$ in the $W$-orbit of the partition
$\mu^{+}\in P_+$ are obtained from $h_{\mu^+}^{(A,B)}$ by
\begin{equation}
  h_{\mu}^{(A,B)} = (c_{w_\mu}^{(A,B)})^{-1}S_{w_\mu}^{(A,B)}h_{\mu^+}^{(A,B)},
  \label{eq:relative_Rodrigues_composition}
\end{equation}
where the coefficients of the top terms $c_{w_\mu}^{(A,B)}$ 
are expressed as
\begin{subequations}
  \begin{eqnarray}
    c_{w_\mu}^{(A)} \Define \prod_{\alpha\in R_{w_\mu}}
    \frac{\langle\alpha^{\vee},
        \mu^+ + a\rho\rangle^2-a^{2}}{\langle\alpha^{\vee},
        \mu^+ + a\rho\rangle},\\
    c_{w_\mu}^{(B)} \Define \prod_{\alpha\in R_{w_\mu}}
    \frac{\langle\alpha^{\vee},
        \mu^+ + 2a\rho\rangle^2-a^{2}
        \bigl(1+(-1)^{\langle\alpha^{\vee},
        \mu\rangle}\bigr)^{2}}{\langle\alpha^{\vee},
        \mu^+ + 2a\rho\rangle}.
  \end{eqnarray}
  \label{eq:relative_coeff_composition}
\end{subequations}
\label{pr:relative_Rodrigues_composition}
\end{proposition}
\Proof The proposition is verified by 
\eref{eq:S-e-d-prop} and \eref{eq:S_expansions} for
$\langle\alpha_{j}^{\vee},\mu\rangle>0$. \QED 

Combining Propositions \ref{pr:Rodrigues_partition} and 
\ref{pr:relative_Rodrigues_composition}, 
we immediately obtain the Rodrigues formula
for the non-symmetric multivariable Hermite and Laguerre
polynomials with a general composition $h_{\mu}^{(A,B)}$, $\mu\in P$.
\begin{theorem}[Rodrigues formula] The monic non-symmetric multivariable
Hermite and Laguerre polynomials $h_{\mu}^{(A,B)}$
with a general composition $\mu\in P$ are algebraically obtained by
applying the raising operators $A_{\mu^+}^{(A,B)\dagger}$ and
the product of braid operators $S_{w_\mu}$ to $h_0^{(A,B)} = 1$,
\begin{equation}
  h_{\mu}^{(A,B)}=(c_{w_\mu}^{(A,B)}c_{\mu^+}^{(A,B)})^{-1}
  S_{w_\mu}^{(A,B)}A_{\mu^+}^{(A,B)\dagger}h_0^{(A,B)}.
  \label{eq:Rodrigues_generic}
\end{equation}
\label{th:Rodrigues_generic}
\end{theorem}
We note that the corresponding formulas for the eigenstates,
$\varphi_\mu^{(A,B)}(x)\Define h_\mu^{(A,B)}(x)\phi_{\rm g}^{(A,B)}(x),
\quad \mu\in P$, of the original Hamiltonian
$\hat{\mathcal H}^{(A,B)}$ \eref{eq:Calogero_Hamiltonians} are
\begin{eqnarray*}
  \varphi_{\mu}^{(A,B)} & = \phi_{\rm g}^{(A,B)}
  (c_{w_\mu}^{(A,B)}c_{\mu^+}^{(A,B)})^{-1}
  S_{w_\mu}^{(A,B)}A_{\mu^+}^{(A,B)\dagger}h_0^{(A,B)} \\
  & = (c_{w_\mu}^{(A,B)}c_{\mu^+}^{(A,B)})^{-1}
  \hat{S}_{w_\mu}^{(A,B)}\hat{A}_{\mu^+}^{(A,B)\dagger}\phi_{\rm g}^{(A,B)},
\end{eqnarray*}
where $\hat{S}_{w_\mu}^{(A,B)}\Define\phi_{\rm g}^{(A,B)}\circ
S_{w_\mu}^{(A,B)}\circ(\phi_{\rm g}^{(A,B)})^{-1}$ and
$\hat{A}_{\mu^+}^{(A,B)\dagger}\Define\phi_{\rm g}^{(A,B)}\circ
A_{\mu^+}^{(A,B)\dagger}\circ(\phi_{\rm g}^{(A,B)})^{-1}$.

Now we shall calculate norms of the non-symmetric multivariable
Hermite and Laguerre polynomials in an algebraic fashion
using the Rodrigues formula.
We also use the norms for $h_0^{(A,B)}=1$:
\begin{eqnarray*}
  \langle h_0^{(A)},h_0^{(A)}\rangle_{(A)} 
  =\frac{(2\pi)^{\frac{N}{2}}}
  {(2\omega)^{\frac{1}{2}N(Na+(1-a))}}\prod_{j\in I}
  \frac{\Gamma(1+ja)}{\Gamma(1+a)}, \\
  \langle h_0^{(B)},h_0^{(B)}\rangle_{(B)}
  =\frac{1}{\omega^{N(N-1)a+N(b+\frac{1}{2})}}\prod_{j\in I}
  \frac{\Gamma(1+ja)\Gamma((j-1)a+b+\frac{1}{2})}{\Gamma(1+a)}.
\end{eqnarray*}
which are proved by certain limits of the Selberg integral
\cite{Macdonald_3,Mehta_1}.

First we shall calculate the square norms of the polynomials with a
general composition divided by those with the corresponding partition.
\begin{lemma}[cf.~\cite{Baker_2,Knop_1}]
For the non-symmetric multivariable Hermite and 
Laguerre polynomials with a general composition $h_{\mu}^{(A,B)}$,
$\mu\in W(\mu^{+})$, $\mu^+\in P_+$, we have
\begin{subequations}
  \begin{eqnarray}
    \fl \langle h_{\mu}^{(A)},h_{\mu}^{(A)} \rangle_{(A)} 
    & =  \prod_{\alpha\in R_{w_\mu}}
    \frac{\langle \alpha^{\vee},\mu^+ +a\rho\rangle^2}%
    {\langle \alpha^{\vee},\mu^+ +a\rho\rangle^2 -a^2}
    \langle h_{\mu^+}^{(A)},h_{\mu^+}^{(A)}\rangle, \\
    \fl \langle h_{\mu}^{(B)},h_{\mu}^{(B)} \rangle_{(B)}
    & = \prod_{\alpha\in R_{w_\mu}}
    \frac{\langle \alpha^{\vee},\mu^+ +2a\rho\rangle^2}%
    {\langle \alpha^{\vee},\mu^+ +2a\rho\rangle^2 
    -a^2(1+(-1)^{\langle \alpha^{\vee},\mu^+\rangle})^2}
    \langle h_{\mu^+}^{(B)},h_{\mu^+}^{(B)}\rangle,
  \end{eqnarray}
  \label{eq:relative_norms_for_composition}
\end{subequations}
which are independent of the choice of reduced expressions of $w_{\mu}$.
\label{lm:relative_norms_for_composition}
\end{lemma}
\Proof 
Due to Proposition \ref{pr:relative_Rodrigues_composition},
the square norms of the non-symmetric multivariable Hermite and Laguerre
polynomials with a general composition $h_{\mu}^{(A,B)}$, $\mu\in P$ are
expressed as
\begin{eqnarray*}
  \langle h_{\mu}^{(A,B)},h_{\mu}^{(A,B)} \rangle_{(A,B)}
  & =(c_{w_\mu}^{(A,B)})^{-2}
  \langle S_{w_\mu}^{(A,B)}h_{\mu^+}^{(A,B)}, 
  S_{w_\mu}^{(A,B)}h_{\mu^+}^{(A,B)}\rangle_{(A,B)}\\
  & = (c_{w_\mu}^{(A,B)})^{-2}
  \langle h_{\mu^+}^{(A,B)}, S_{w_\mu}^{(A,B)\dagger}
  S_{w_\mu}^{(A,B)}h_{\mu^+}^{(A,B)}\rangle_{(A,B)},
\end{eqnarray*}
where $S_{w_\mu}^{(A,B)\dagger} = (-S_{j_1})(-S_{j_2})\cdots(-S_{j_l})$.
Using \eref{eq:S-e-prop}, we have
\begin{eqnarray*}
  \fl \langle h_{\mu}^{(A)},h_{\mu}^{(A)} \rangle_{(A)}
  =(c_{w_\mu}^{(A)})^{-2}\prod_{n=1}^{l}
  \bigl(\langle \alpha^{\vee}_{j_n},\mu^{(n-1)}+a\rho(\mu^{(n-1)})
  \rangle^2 -a^2\bigr)
  \langle h_{\mu^+}^{(A)},h_{\mu^+}^{(A)}\rangle_{(A)},\\
  \fl \langle h_{\mu}^{(B)},h_{\mu}^{(B)} \rangle_{(B)}
  =(c_{w_\mu}^{(B)})^{-2}\prod_{n=1}^{l}
  \bigl(\langle \alpha^{\vee}_{j_n},\mu^{(n-1)}+2a\rho(\mu^{(n-1)})\rangle^2 
  -a^2(1+(-1)^{\langle \alpha^{\vee}_{j_n},\mu^{(n-1)}\rangle})^2\bigr)
  \langle h_{\mu^+}^{(B)},h_{\mu^+}^{(B)}\rangle_{(B)},
\end{eqnarray*}
where the sequence of compositions $\{\mu^{(n)}|n=1,2,\cdots l\}$ is
defined by a reduced expression of $w_{\mu}$ as 
in \eref{eq:ordered_sequence}.

To proceed calculation, we need a property related to the reflection.
For any $s_j\in W$ and $\mu\in P$ such that 
$\mu\neq s_j(\mu)$, the following formula holds:
\begin{equation}
  s_j(\rho(\mu))=\rho(s_j(\mu)),
  \label{eq:lemma_rho}
\end{equation}
because $w_{s_j(\mu)}=s_j w_{\mu}$, if $\mu\neq s_j(\mu)$.
From the relations above, we can easily verify
\begin{eqnarray*}
 \label{eq:lemma_sequence}
  \langle \alpha^{\vee}_{j_n},\mu^{(n-1)}\rangle = 
  \langle s_{j_n}(\alpha^{\vee}_{j_n}),\mu^{(n-2)}\rangle
  =\cdots
  =\langle s_{j_1}s_{j_2}\cdots s_{j_{n-1}}(\alpha^{\vee}_{j_n}),
    \mu^+\rangle
  \\
  \langle \alpha^{\vee}_{j_n},\rho(\mu^{(n-1)})\rangle=
  \langle s_{j_n}(\alpha^{\vee}_{j_n}),\rho(\mu^{(n-2)})\rangle
  =\cdots
  =\langle s_{j_1}s_{j_2}\cdots s_{j_{n-1}}
  (\alpha^{\vee}_{j_n}),\rho \rangle.
\end{eqnarray*}
Here we have used $\rho(\mu^+)=\rho$, $\mu^+\in P_+$, which 
follows from the definition of $\rho(\mu)$, $\mu\in P$. Since the set of roots
$\{s_{j_1}s_{j_2}\cdots s_{j_{n-1}}(\alpha_{j_n})|n=1,2,\cdots,l\}$
is nothing but $R_{w_\mu}$ that is uniquely determined by choosing $\mu\in P$,
the square norms $\langle h_{\mu}^{(A,B)},h_{\mu}^{(A,B)} \rangle_{(A,B)}$ can
be expressed as
\begin{eqnarray*}
  \fl \langle h_{\mu}^{(A)},h_{\mu}^{(A)} \rangle_{(A)} 
  & = (c_{w_\mu}^{(A)})^{-2}\prod_{\alpha\in R_{w_\mu}}
  \bigl(\langle \alpha^{\vee},\mu^+ +a\rho\rangle^2 -a^2\bigr)
  \langle h_{\mu^+}^{(A)},h_{\mu^+}^{(A)}\rangle_{(A)} \nnn
  & =  \prod_{\alpha\in R_{w_\mu}}
  \frac{\langle \alpha^{\vee},\mu^+ +a\rho\rangle^2}%
  {\langle \alpha^{\vee},\mu^+ +a\rho\rangle^2 -a^2}
  \langle h_{\mu^+}^{(A)},h_{\mu^+}^{(A)}\rangle_{(A)}, \\
  \fl \langle h_{\mu}^{(B)},h_{\mu}^{(B)} \rangle_{(B)}
  & = (c_{w_\mu}^{(B)})^{-2}\prod_{\alpha\in R_{w_\mu}}
  \bigl(\langle \alpha^{\vee},\mu^+ +2a\rho\rangle^2 
  -a^2(1+(-1)^{\langle \alpha^{\vee},\mu^+\rangle} )^2\bigr)
  \langle h_{\mu^+}^{(B)},h_{\mu^+}^{(B)}\rangle_{(B)} \nnn
  & = \prod_{\alpha\in R_{w_\mu}}
  \frac{\langle \alpha^{\vee},\mu^+ +2a\rho\rangle^2}%
  {\langle \alpha^{\vee},\mu^+ +2a\rho\rangle^2 
  -a^2(1+(-1)^{\langle \alpha^{\vee},\mu^+\rangle})^2}
  \langle h_{\mu^+}^{(B)},h_{\mu^+}^{(B)}\rangle_{(B)},
\end{eqnarray*}
which are nothing but the formulas of Lemma
\ref{lm:relative_norms_for_composition}. \QED

The square norms of the polynomials with a partition are summarized
as follows.
\begin{proposition}
The square norms of the non-symmetric multivariable Hermite and 
Laguerre polynomials $h_{\mu}^{(A,B)}$ with a partition
$\mu\in P_+$ are given by
\begin{subequations}
  \begin{eqnarray}
    \fl\langle h_{\mu}^{(A)},h_{\mu}^{(A)} \rangle_{(A)} \nnn
    \fl\quad = \frac{(2\pi)^{\frac{N}{2}}}{(2\omega)^{\frac{1}{2}N(Na+(1-a))%
    +|\mu|}}
    \prod_{i\in I}\Gamma\bigl(\mu_{i}+a(N-i)+1\bigr)\nnn
    \fl\qquad \prod_{\alpha\in R_{+}}
    \frac{\Gamma\bigl(\langle\alpha^{\vee},\mu+a\rho\rangle+1+a\bigr)
    \Gamma\bigl(\langle\alpha^{\vee},\mu+a\rho\rangle+1-a\bigr)}
    {\Gamma\bigl(\langle\alpha^{\vee},\mu+a\rho\rangle+1\bigr)^2},\\
    \fl\langle h_{\mu}^{(B)},h_{\mu}^{(B)} \rangle_{(B)} \nnn
    \fl\quad = \frac{1}{\omega^{N(N-1)a+N(b+\frac{1}{2})+|\mu|}}
    \prod_{i\in I}
    \Gamma\bigl(\bigl[\frac{1}{2}(\mu_{i}+1)\bigr]
    +a(N-i)+b+\frac{1}{2}\bigr)
    \Gamma\bigl(\bigl[\frac{1}{2}\mu_{i}\bigr]+a(N-i)+1\bigr)\nnn
    \fl\qquad \prod_{\alpha\in R_+}
    \frac{\Gamma\bigl(\bigl[\frac{1}{2}\langle\alpha^{\vee},\mu\rangle\bigr]
    +\langle\alpha^{\vee},a\rho\rangle+1+a\bigr)
    \Gamma\bigl(\bigl[\frac{1}{2}\langle\alpha^{\vee},\mu\rangle\bigr]
    +\langle\alpha^{\vee},a\rho\rangle+1-a\bigr)}
    {\Gamma\bigl(\bigl[\frac{1}{2}\langle\alpha^{\vee},\mu\rangle\bigr]
    +\langle\alpha^{\vee},a\rho\rangle+1\bigr)^2},\nnn
  \end{eqnarray}
  \label{eq:norms_for_partition}
\end{subequations}
where the Gauss's symbol $[x]$ means the maximum integer 
that is less than $x$.
\label{pr:norms_for_partition}
\end{proposition}
\Proof
In order to calculate the square norms of the polynomials with 
partitions $\mu\in P_+$ algebraically, we use the following relations
\begin{eqnarray*}
  \fl A_{j}^{(A)}A_{j}^{(A)\dagger}=\frac{1}{2\omega}
  \prod_{k=1}^{j}(d_{k}^{(A)}+1)
  \prod_{l=j+1}^{N}\bigl((d_{k}^{(A)}-d_{l}^{(A)}+1)^2 - a^2\bigr),\\
  \fl A_{j}^{(B)}A_{j}^{(B)\dagger}=\frac{1}{2\omega}
  \prod_{k=1}^{j}(d_{k}^{(B)}+bt_k +1)
  \prod_{l=j+1}^{N}\bigl((d_{k}^{(B)}-d_{l}^{(B)}+1)^2 
  - 2a^2(1-t_k t_l)\bigr),
\end{eqnarray*}
which can be verified by \eref{eq:S-e-prop}, \eref{eq:S-e-d-prop}
and \eref{eq:raising_operators}, and the defining
relations of the non-symmetric multivariable Hermite and Laguerre polynomials
\eref{eq:non-symmetric_polynomials}. Then we have
\begin{eqnarray*}
  \fl\langle h_{\mu}^{(A)},h_{\mu}^{(A)} \rangle_{(A)} \nnn
  \fl\quad = (2\omega)^{-|\mu|}(c_{\mu}^{(A)})^{-2}\prod_{i=1}^N \prod_{j=1}^i 
  \prod_{k=1}^{\mu_{i}-\mu_{i+1}}\bigl(\mu_{i}-k+1+a(N-j)\bigr)\nnn
  \fl \qquad
     \prod_{m=i+1}^{N}\bigl((\mu_{i}-\mu_{m}-k+1+a(m-j))^2 -a^2\bigr)
     \langle h_{0}^{(A)},h_{0}^{(A)} \rangle_{(A)}\nnn
  \fl\quad 
  = (2\omega)^{-|\mu|}
  (c_{\mu}^{(A)})^{-2}\langle h_{0}^{(A)},h_{0}^{(A)} \rangle_{(A)}
  \prod_{i\in I}\prod_{j=1}^{\mu_{i}}\bigl(\mu_{i}-j+1+a(N-i)\bigr)\nnn
  \fl\qquad \prod_{\alpha\in R_+}\Xmae\prod_{k=1}
  ^{\langle \alpha^{\vee},\mu\rangle}
  \Xmae\bigl(\langle\alpha^{\vee},\mu+a\rho\rangle -k+1 +a\bigr)
  \bigl(\langle\alpha^{\vee},\mu+a\rho\rangle -k+1 -a\bigr)\nnn
  \fl\quad = \frac{(2\pi)^{\frac{N}{2}}}{(2\omega)^{\frac{1}{2}N(Na+(1-a))%
  +|\mu|}}
  \prod_{i\in I}\Gamma\bigl(\mu_{i}+a(N-i) +1\bigr)\nnn
  \fl\qquad \prod_{\alpha\in R_+}
  \frac{\Gamma\bigl(\langle\alpha^{\vee},\mu+a\rho\rangle +1 +a\bigr)
  \Gamma\bigl(\langle\alpha^{\vee},\mu+a\rho\rangle +1 -a\bigr)}
  {\Gamma\bigl(\langle\alpha^{\vee},\mu+a\rho\rangle +1 \bigr)^2},\\
  \fl\langle h_{\mu}^{(B)},h_{\mu}^{(B)} \rangle_{(B)} \nnn
  \fl\quad = (2\omega)^{-|\mu|}(c_{\mu}^{(B)})^{-2}\prod_{i=1}^N \prod_{j=1}^i 
  \prod_{k=1}^{\mu_{i}-\mu_{i+1}}\bigl(\mu_{i}-k+1+2a(N-j)
  +b(1-(-1)^{\mu_{i}-k+1})\bigr)\nnn
  \fl\qquad \prod_{m=i+1}^{N}\bigl((\mu_{i}-\mu_{m}-k+1+2a(m-j))^2 
  -2a^2(1+(-1)^{\mu_{i}- \mu_{m}- k+1})\bigr)
     \langle h_{0}^{(B)},h_{0}^{(B)} \rangle_{(B)}\nnn
  \fl\quad = (2\omega)^{-|\mu|}
  (c_{\mu}^{(B)})^{-2}\langle h_{0}^{(B)},h_{0}^{(B)} \rangle_{(B)}
  \prod_{i\in I}\prod_{j=1}^{\mu_{i}}\bigl(\mu_{i}-j+1+2a(N-i)
  +b(1-(-1)^{\mu_{i}-j+1})\bigr)\nnn
  \fl\qquad \prod_{\alpha\in R_+}\Xmae\prod_{k=1}%
  ^{\langle \alpha^{\vee},\mu\rangle}
  \Xmae\bigl(\langle\alpha^{\vee},\mu+2a\rho\rangle -k+1 
  +a(1+(-1)^{\langle\alpha^{\vee},\mu\rangle -k+1})\bigr)\nnn
  \fl \qquad\qquad\DMusr\Xusr
  \bigl(\langle\alpha^{\vee},\mu+2a\rho\rangle -k+1 
  -a(1+(-1)^{\langle\alpha^{\vee},\mu\rangle -k+1})\bigr)\nnn
  \fl\quad = \frac{1}{\omega^{N(N-1)a+N(b+\frac{1}{2})+|\mu|}}
  \prod_{i\in I}
  \Gamma\bigl(\bigl[\frac{1}{2}(\mu_{i}+1)\bigr]+a(N-i)+b+\frac{1}{2}\bigr)
  \Gamma\bigl(\bigl[\frac{1}{2}\mu_{i}\bigr]+a(N-i)+1\bigr)\nnn
  \fl\qquad \prod_{\alpha\in R_+}
  \frac{\Gamma\bigl(\bigl[\frac{1}{2}\langle\alpha^{\vee},\mu\rangle\bigr]
  +\langle\alpha^{\vee},a\rho\rangle +1 +a\bigr)
  \Gamma\bigl(\bigl[\frac{1}{2}\langle\alpha^{\vee},\mu\rangle\bigr]
  +\langle\alpha^{\vee},a\rho\rangle +1 -a\bigr)}%
  {\Gamma\bigl(\bigl[\frac{1}{2}\langle\alpha^{\vee},\mu\rangle\bigr]
  +\langle\alpha^{\vee},a\rho\rangle +1 \bigr)^2},\nnn
\end{eqnarray*}
which prove Proposition \ref{pr:norms_for_partition}. \QED

Finally, we obtain the formulas for the square norms of the
non-symmetric multivariable Hermite and Laguerre polynomials with
a general composition including orthogonality.
\begin{theorem}
For the non-symmetric multivariable Hermite and Laguerre
polynomials $h_{\mu}^{(A,B)}$ with a general composition 
$\mu\in W(\mu^{+})$, $\mu^+\in P_+$, we have
\begin{subequations}
  \begin{eqnarray}
    \fl\langle h_{\mu}^{(A)},h_{\nu}^{(A)} \rangle_{(A)} \nnn
    \fl\quad = \delta_{\mu,\nu}
    \frac{(2\pi)^{\frac{N}{2}}}{(2\omega)^{\frac{1}{2}N(Na+(1-a))+|\mu|}}
    \prod_{\beta\in R_{w_\mu}}
    \frac{\langle \beta^{\vee},\mu^+ +a\rho\rangle^2}%
    {\langle \beta^{\vee},\mu^{+}+a\rho\rangle^2 -a^2}
    \prod_{i\in I}\Gamma\bigl(\mu_{i}^{+}+a(N-i) +1\bigr)\nnn
    \fl\qquad\prod_{\alpha\in R_+}
    \frac{\Gamma\bigl(\langle\alpha^{\vee},\mu^{+}+a\rho\rangle+1+a\bigr)
    \Gamma\bigl(\langle\alpha^{\vee},\mu^{+}+a\rho\rangle+1-a\bigr)}%
    {\Gamma\bigl(\langle\alpha^{\vee},\mu^{+}+a\rho\rangle +1 \bigr)^2},\\
    \fl\langle h_{\mu}^{(B)},h_{\nu}^{(B)} \rangle_{(B)} \nnn
    \fl\quad = \delta_{\mu,\nu}
    \frac{1}{\omega^{N(N-1)a+N(b+\frac{1}{2})+|\mu|}}
    \prod_{\beta\in R_{w_\mu}}
    \frac{\langle \beta^{\vee},\mu^{+}+2a\rho\rangle^2}%
    {\langle \beta^{\vee},\mu^{+} +2a\rho\rangle^2 
    -a^2(1+(-1)^{\langle \beta^{\vee},\mu^{+}\rangle})^2}\nnn
    \fl\qquad\Xusr\prod_{i\in I}
    \Gamma\bigl(\bigl[\frac{1}{2}(\mu_{i}^{+}+1)\bigr]
    +a(N-i)+b+\frac{1}{2}\bigr)
    \Gamma\bigl(\bigl[\frac{1}{2}\mu_{i}^{+}\bigr]+a(N-i)+1\bigr)\nnn
    \fl\qquad\prod_{\alpha\in R_+}
    \frac{\Gamma\bigl(\bigl[\frac{1}{2}\langle\alpha^{\vee},
    \mu^{+}\rangle\bigr]
    +\langle\alpha^{\vee},a\rho\rangle+1+a\bigr)
    \Gamma\bigl(\bigl[\frac{1}{2}\langle\alpha^{\vee},\mu^{+}\rangle\bigr]
    +\langle\alpha^{\vee},a\rho\rangle+1-a\bigr)}%
    {\Gamma\bigl(\bigl[\frac{1}{2}\langle\alpha^{\vee},\mu^{+}\rangle\bigr]
    +\langle\alpha^{\vee},a\rho\rangle+1 \bigr)^2}.\nnn
  \end{eqnarray}
  \label{eq:norms_for_generic_weight}
\end{subequations}
\label{pr:norms_for_generic_weight}
\end{theorem}
In terms of the eigenstates of the original Calogero Hamiltonian,
the above orthogonality relations are expressed by
$(\varphi_{\mu}^{(A,B)},\varphi_{\nu}^{(A,B)})
=\langle h_{\mu}^{(A,B)},h_{\nu}^{(A,B)} \rangle_{(A,B)}$.

Thus we have presented an algebraic method that enables us to
obtain all the non-symmetric 
multivariable Hermite and Laguerre polynomials with general compositions
and their square norms. 

\section{Symmetrization and anti-symmetrization}
\label{sec:(anti-)symmetrization}

We can
readily confirm that the non-symmetric multivariable Hermite and Laguerre
polynomials with compositions $\mu$ lying in the same $W$-orbit of the
partition $\mu^{+}$ have the same eigenvalue of the 
Hamiltonians \eref{eq:transformed_Hamiltonian},
\[
  {\mathcal H}^{(A,B)}h_{\mu}^{(A,B)}=\omega|\mu^{+}|h_{\mu}^{(A,B)},
  \quad\mbox{ for }\, \mu\in W(\mu^+),\quad \mu^+\in P_+ .
\]
More generally, the polynomials with compositions $\mu\in W(\mu^+)$ have
the same eigenvalue of an arbitrary symmetric polynomial, e.g.,
any of the power sums, of the Cherednik operators. 
Thus any linear combinations of $h_{\mu}^{(A,B)}$, 
$\mu\in W(\mu^+)$, $\mu^+\in P_+$ are eigenfunctions of the Calogero 
Hamiltonians ${\mathcal H}^{(A,B)}$ and all of their higher order
conserved operators. 

Among all such linear combinations,
we shall deal with symmetric and anti-symmetric eigenvectors of the
Calogero Hamiltonians in ${\mathbb C}[x]^{\pm W}$ that respectively
correspond to the bosonic and the fermionic eigenstates of the models.
We symmetrize and anti-symmetrize non-symmetric eigenvectors, but
our formulation does not use the symmetrizer or the anti-symmetrizer 
\cite{Baker_1} that make coefficients of top terms different
from unity. 
To describe the anti-symmetric eigenvectors,
we introduce sublattices of $P_{+}$ 
such as $P_{+}+\delta\Define\{\mu+\delta|\mu\in P_{+} \}$ and so forth.
Other sublattices of $P_+$ in what follows are defined in a similar way.
We notice that, for the $B_N$-case,
the parity with respect to each variable is restricted
to even or odd 
since the symmetric and anti-symmetric eigenvectors are
eigenvectors of the reflection operators $\{t_{j}|j\in I\}$ at the same time.

\begin{theorem}
Let $H_{\mu^{+}}^{(A,B)+}, (\mu^{+}\in P_{+})$,
    $H_{\mu^{+}}^{(A)-}, (\mu^{+}\in P_{+}+\delta)$
and $H_{\mu^{+}}^{(B)-}, (\mu^{+}\in P_{+}+2\delta)$ be
the following linear combinations of the corresponding non-symmetric 
polynomials with compositions $\mu\in W(\mu^{+})$:
\begin{equation}
  H_{\mu^{+}}^{(A,B)\pm}=\sum_{\mu\in W(\mu^+)}
  b_{\mu^+\mu}^{(A,B)\pm}h_{\mu}^{(A,B)},
  \label{eq:linear_combinations}
\end{equation}
whose coefficients are 
\begin{equation}
   \fl b_{\mu^+\mu}^{(A)\pm}=\prod_{\alpha\in R_{w_\mu}}
    \pm\frac{\langle\alpha,\mu^+ + a\rho\rangle \mp a}%
    {\langle\alpha,\mu^+ + a\rho\rangle}, \quad 
    b_{\mu^+\mu}^{(B)\pm}=\prod_{\alpha\in R_{w_\mu}}
    \pm\frac{\langle\alpha,\mu^+ + 2a\rho\rangle \mp 2a}%
    {\langle\alpha,\mu^+ + 2a\rho\rangle}.
  \label{eq:(anti-)symmetrizing_coefficients}
\end{equation}
Then we find $H_{\mu^+}^{(A,B)\pm}\in\mathbb{C}[x]^{\pm W}$,
which we call symmetric and anti-symmetric 
multivariable Hermite and Laguerre polynomials, respectively.
\label{prop:(anti-)symmetrization}
\end{theorem}
\Proof
We consider $H_{\mu^+}^{(A,B)\pm}$ 
of the forms (\ref{eq:linear_combinations}).
By requiring $H_{\mu^+}^{(A,B)\pm}\in\mathbb{C}[x]^{\pm W}$
and $b_{\mu^+\mu^+}^{(A,B)\pm}=1$, which is clearly equivalent to the
requirement that the coefficients of the top symmetrized monomial are unity,
the coefficients $b_{\mu^+\mu}^{(A,B)\pm}$ are uniquely determined.
The proofs of the above two formulas are almost the same and we shall
only show a proof for $b_{\mu^+\mu}^{(B)\pm}$.
Note that, for the $B_N$-case,
$\mu^{+}\in 2P_{+}$ (all even) 
or $\mu^{+}\in 2P_{+}+1^N$ (all odd)
so that the parities of all the variables are the same.
From Lemma \ref{lm:K_expansions}, we have
\[
  b_{\mu^+\mu}^{(B)\pm}h_{\mu}^{(B)}+b_{\mu^+s_j(\mu)}^{(B)\pm}
  h_{s_j(\mu)}^{(B)}
  = \pm K_j\Bigl(
  b_{\mu^+\mu}^{(B)\pm}h_{\mu}^{(B)}+b_{\mu^+s_j(\mu)}^{(B)\pm}
  h_{s_j(\mu)}^{(B)}\Bigr).
\]
Without loss of generality, we may assume $\mu\succ s_j(\mu)$
(i.e.~$\langle\alpha_j^{\vee},\mu\rangle > 0$) 
since the case $\mu = s_j(\mu)$ 
(i.e.~$\langle\alpha_j^{\vee},\mu\rangle = 0$) is trivial.
Then the above relation is rewritten as
\begin{eqnarray*}
  \fl \lefteqn{
    b_{\mu^+\mu}^{(B)\pm}h_{\mu}^{(B)}
    +b_{\mu^+s_j(\mu)}^{(B)\pm}h_{s_j(\mu)}^{(B)}
  }\nnn
  \fl\quad = \pm\biggl(
    & \Bigl(
    \frac{2a}{\langle\alpha_j^{\vee},\mu+2a\rho(\mu)\rangle}
    b_{\mu^+\mu}^{(B)\pm} + b_{\mu^+s_j(\mu)}^{(B)\pm}\Bigr)h_{\mu}^{(B)}\nnn
    \fl & +\Bigl(\frac{\langle\alpha_j^{\vee},\mu+2a\rho(\mu)\rangle^2-4a^2}%
    {\langle\alpha_j^{\vee},\mu+2a\rho(\mu)\rangle^2}b_{\mu^+\mu}^{(B)\pm}
    -\frac{2a}{\langle\alpha_j^{\vee},\mu+2a\rho(\mu)\rangle}
    b_{\mu^+s_j(\mu)}^{(B)\pm}\Bigr)h_{s_j(\mu)}^{(B)}\biggr),
\end{eqnarray*}
where we have used $1+(-1)^{\langle\alpha_j^{\vee},\mu\rangle}=2$ for
$\mu\in W(\mu^{+})$
with $\mu^{+}\in 2P_{+}$ or $\mu^{+}\in 2P_{+}+1^N$.
Thus we obtain
\begin{equation}
  \frac{b_{\mu^+s_j(\mu)}^{(B)\pm}}{b_{\mu^+\mu}^{(B)\pm}}
  =\pm\frac{\langle\alpha_j^{\vee},\mu+2a\rho(\mu)\rangle \mp 2a}%
    {\langle\alpha_j^{\vee},\mu+2a\rho(\mu)\rangle},
  \quad \mbox{for }\mu\prec s_j(\mu).
  \label{eq:recursion_coeff}
\end{equation}
Let $s_{j_l}\cdots s_{j_2}s_{j_1}$ and 
$\{\mu^{(n)}\in W(\mu^{+})|n=1,2,\cdots,l\}$ 
be a reduced expression of $w_{\mu}$ and
a sequence of compositions as have been given in \eref{eq:ordered_sequence}.
Iterated use of the recursion relation of $b_{\mu^+\mu}^{(B)\pm}$
\eref{eq:recursion_coeff} yields
\begin{eqnarray*}
  b_{\mu^+\mu}^{(B)\pm} & = & b_{\mu^+\mu^+}^{(B)\pm}\prod_{n=1}^{l}
  \frac{b_{\mu^+\mu^{(n)}}^{(B)\pm}}{b_{\mu^+\mu^{(n-1)}}^{(B)\pm}}
  =\prod_{n=1}^{l}
  \pm\frac{\langle\alpha_{j_n}^{\vee},
    \mu^{(n-1)}+2a\rho(\mu^{(n-1)})\rangle \mp 2a}%
    {\langle\alpha_{j_n}^{\vee},\mu^{(n-1)}+2a\rho(\mu^{(n-1)})\rangle},
\end{eqnarray*}
where $b_{\mu^+\mu^+}^{(B)\pm}=1$ should be noted.
By use of equation \eref{eq:lemma_sequence},
the above formula is cast into the following form:
\begin{equation}
 \label{eq:bmumu}
  b_{\mu^+\mu}^{(B)\pm}=\prod_{n=1}^{l}
  \pm\frac{\langle s_{j_1}s_{j_2}\cdots s_{j_{n-1}}
    (\alpha_{j_n}^{\vee}),\mu^+ +2a\rho\rangle \mp 2a}%
    {\langle s_{j_1}s_{j_2}\cdots s_{j_{n-1}}
    (\alpha_{j_n}^{\vee}),\mu^+ +2a\rho\rangle}.
\end{equation}
Recalling the fact $R_{w_{\mu}}=
\{s_{j_1}s_{j_2}\cdots s_{j_{n-1}}(\alpha_{j_n})|n=1,2,\cdots,l\}$, we
verify that (\ref{eq:bmumu}) is nothing but $b_{\mu^+\mu}^{(B)\pm}$ 
in the theorem above. \QED

The symmetric multivariable Hermite and Laguerre polynomials are the same as
those discussed in \cite{Ujino_2,Ujino_1,Ujino_3,Ujino_4,Kakei_1,Baker_0,%
vanDiejen_1}.
There are several equivalent conditions to characterize these symmetric
polynomials. For instance, triangularity in $\mathbb{C}[x]^{W}$
and orthogonality 
with respect to $\langle\cdot,\cdot\rangle_{(A,B)}$ characterize
the symmetric multivariable Hermite and Laguerre polynomials
up to a constant factor. However,
we have implicitly taken another way of characterization.
Those polynomials are identified by
polynomial parts of eigenstates
for all conserved operators of the Calogero models with bosonic particles.
We note 
\begin{equation}
\eqalign{
    K_j H_{\mu}^{(A,B)\pm}
    =\pm H_{\mu}^{(A,B)\pm}, \\
    t_j H_{\mu}^{(B)\pm}
    =\cases{
       H_{\mu}^{(B)\pm} & for $\mu\in 2P_{+}$, \\
      -H_{\mu}^{(B)\pm} & for $\mu\in 2P_{+}+1^N$.
      } 
  } 
\end{equation}
Therefore,
on such (anti-)symmetric functions (with all even or all odd parity for the
$B_N$-case) multiplied by the reference states $\phi_{\rm g}^{(A,B)}$,
the Calogero Hamiltonians \eref{eq:Calogero_Hamiltonians} with distinguishable
particles reduce to
\begin{eqnarray*}
  \fl \hat{\mathcal H}^{(A)\pm}(a) = \frac{1}{2}\sum_{j=1}^{N}
    \bigl(-\frac{\partial^2}{\partial x_j^2}+\omega^{2}x_{j}^{2}\bigr)
    +\frac{1}{2}\sum_{\stackrel{\scriptstyle j,k=1}{j\neq k}}^{N}
    \frac{a^{2}\mp a}{(x_{j}-x_{k})^{2}},\\
  \fl \hat{\mathcal H}^{(B)\pm,\pm}(a,b) = \frac{1}{2}\sum_{j=1}^{N}
    \bigl(-\frac{\partial^2}{\partial x_j^{2}}
    +\omega^{2}x_{j}^{2}+\frac{b^2\mp b}{x_j^2}\bigr) 
    {}+\frac{1}{2}\sum_{\stackrel{\scriptstyle j,k=1}{j\neq k}}^{N}
    \Bigl(\frac{a^{2}\mp a}{(x_{j}-x_{k})^{2}}
    +\frac{a^{2}\mp a}{(x_{j}+x_{k})^{2}}\Bigr),
\end{eqnarray*}
which are the Calogero models with indistinguishable (bosonic or 
fermionic) particles.
The first superscripts $\pm$ on the Hamiltonians correspond
to the double sign $\mp$ before the coupling parameter $a$ that
respectively mean symmetric and anti-symmetric cases.
The second superscripts $\pm$ on the $B_N$-Hamiltonian correspond
to the double sign $\mp$ before the coupling parameter $b$ that
respectively mean all even and all odd cases.
The Calogero models $\hat{\mathcal{H}}^{(A)\pm}(a)$ and
$\hat{\mathcal{H}}^{(B)\pm,\pm}(a,b)$ are diagonalized by
$\Phi_{\mu}^{(A,B)\pm}(x)\Define 
H_{\mu}^{(A,B)\pm}(x)\phi_{\rm g}^{(A,B)}(x)\in
\mathbb{C}[x]^{\pm W}\phi_{\rm g}^{(A,B)}$,
\begin{eqnarray*}
  \hat{\mathcal H}^{(A)\pm}(a)\Phi_{\mu}^{(A)\pm}
  =(\omega|\mu|+E_{\rm g}^{(A)})\Phi_{\mu}^{(A)\pm},\\
  \hat{\mathcal H}^{(B)\pm,\pm}(a,b)\Phi_{\mu}^{(B)\pm}
  =(\omega|\mu|+E_{\rm g}^{(B)})\Phi_{\mu}^{(B)\pm},
\end{eqnarray*}
where the partitions $\mu$ should be chosen from the appropriate sublattices
of $P_+$ corresponding to symmetries and parties.

The above Hamiltonians with indistinguishable particles are related with
each other by the following shifts of the parameters $a$, $b\in
\mathbb{R}_{\geq 0}$,
\begin{eqnarray*}
  \fl\hat{\mathcal H}^{(A)+}(a+1)
  =\hat{\mathcal H}^{(A)-}(a),\\
  \fl\hat{\mathcal H}^{(B)+,+}(a+1,b+1)
  =\hat{\mathcal H}^{(B)-,+}(a,b+1)
  =\hat{\mathcal H}^{(B)+,-}(a+1,b)
  =\hat{\mathcal H}^{(B)-,-}(a,b).
\end{eqnarray*}
Thus each Hamiltonian with indistinguishable particles has both bosonic
and fermionic eigenstates with the appropriate shifts of the parameters 
in the eigenstates.
Directing our attention to polynomial parts of the eigenstates,
we notice that the symmetric and anti-symmetric polynomials 
(with all even or all odd parties for the Laguerre case) are mutually 
related by
\begin{eqnarray*}
  \Delta(x)H_{\mu}^{(A)+}(x;a+1)
  =H_{\mu+\delta}^{(A)-}(x;a), \quad \mu \in P_+,\\
  \Delta_1(x)\Delta_2(x) H_{\mu}^{(B)+}(x;a+1,b+1)
  =\Delta_2(x)H_{\mu+2\delta}^{(B)-}(x;a,b+1) \\
  =\Delta_1(x)H_{\mu+1^N}^{(B)+}(x;a+1,b) 
  =H_{\mu+2\delta+1^N}^{(B)-}(x;a,b), \quad \mu\in 2P_+,
\end{eqnarray*}
where $\Delta(x)\Define\prod_{\stackrel{\scriptstyle i,j\in I}{i<j}}(x_i-x_j)$,
$\Delta_1(x)\Define\prod_{\stackrel{\scriptstyle i,j\in I}{i<j}}(x_i^2-x_j^2)$ 
and $\Delta_2(x)\Define\prod_{i\in I}x_i$.
These results for the symmetric or anti-symmetric multivariable Hermite
and Laguerre polynomials correspond to those for the Jack and
the Macdolald polynomials~\cite{Cherednik_4}.
The above results means that 
the difference of the bosonic and fermionic eigenstates
of the Hamiltonians 
$\hat{\mathcal{H}}^{(A)\pm}(a)$ and $\hat{\mathcal{H}}^{(B)\pm,\pm}(a,b)$,
i.e., products of the reference states and the polynomial parts, comes
from the difference of choice of the sign of 
the difference products, $\Delta(x)$ and $\Delta_1(x)$. Similarly,
the difference of the parity of the eigenstates of
$\hat{\mathcal{H}}^{(B)\pm,\pm}(a,b)$ comes from that of choice of the
sign of $\Delta_2(x)$. To be brief, the statistics of the indistinguishable
particles (the symmetry of the eigenstates) and the parity of the eigenstates
of the Calogero Hamiltonians,
$\hat{\mathcal{H}}^{(A)\pm}(a)$ and $\hat{\mathcal{H}}^{(B)\pm,\pm}(a,b)$,
are respectively determined only by the choice of the sign of the 
difference products, $\Delta(x)$ and $\Delta_1(x)$, 
and the product $\Delta_2(x)$.
We note that similar shifts of the parameter $a$ among the symmetric 
polynomials are
realized by operation of the shift operators on $H_{\mu}^{(A)+}$ and
$H_{2\mu}^{(B)+}$, $\mu\in P_+$. The shift operators give recursion
relations of the square norms of the symmetric polynomials with respect
to the parameter $a$~\cite{Kakei_1,Opdam_1}.

{}From the square norms of the non-symmetric polynomials $\langle
h_{\mu}^{(A,B)},h_{\mu}^{(A,B)}\rangle_{(A,B)}$ and the 
coefficients $b_{\mu^+ \mu}^{(A,B)\pm}$, we shall evaluate
the square norms of the (anti-)symmetric eigenfunctions.
To prove the formula of the square norms, we need the following lemma.
\begin{lemma}
For $\mu\in P_+$, we have an identity,
\[
  \sum_{\nu\in W(\mu)}\prod_{\alpha\in R_{w_\nu}}
  \frac{\langle\alpha^{\vee},\mu+a\rho\rangle \mp a}
  {\langle\alpha^{\vee},\mu+a\rho\rangle \pm a}
  =N!\!\!\prod_{\alpha\in R_+}\frac{\langle\alpha^{\vee},\mu+a\rho\rangle}
  {\langle\alpha^{\vee},\mu+a\rho\rangle \pm a}.
\]
\label{lm:from_Poincare}
\end{lemma}
The above lemma is proved by use of an expression of the Poincar\'e
polynomials \cite{Macdonald_2,Nishino_5} in \ref{app:from_Poincare}.
\begin{theorem}
Let $\mu\in P_{+}$ for $H_{\mu}^{(A)+}$,
$\mu\in P_{+}+\delta$ for $H_{\mu}^{(A)-}$,
$\mu\in 2P_{+}$ or $2P_{+}+1^N$ for $H_{\mu}^{(B)+}$ and
$\mu\in 2(P_{+}+\delta)$ or $2(P_{+}+\delta)+1^N$
for $H_{\mu}^{(B)-}$.
The square norms of the (anti-)symmetric multivariable Hermite and
Laguerre polynomials including orthogonality are presented by
\begin{subequations}
  \begin{eqnarray}
    \fl\eqalign{
    \lefteqn{\langle H_{\mu}^{(A)\pm},H_{\nu}^{(A)\pm}\rangle_{(A)}}\\
    \Musr = & \ \delta_{\mu,\nu}
    \dfrac{(2\pi)^{\frac{N}{2}}N!}{(2\omega)^{\frac{1}{2}N(Na+(1-a))+|\mu|}}
    \prod_{j\in I}\Gamma\bigl(\mu_j+a(N-j)+1\bigr) \\
    & \prod_{\alpha\in R_+}\frac{%
    \Gamma\bigl(\langle\alpha^\vee,\mu+a\rho\rangle +1\mp a\bigr)%
    \Gamma\bigl(\langle\alpha^\vee,\mu+a\rho\rangle \pm a\bigr)}%
    {\Gamma\bigl(\langle\alpha^\vee,\mu+a\rho\rangle +1\bigr)%
    \Gamma\bigl(\langle\alpha^\vee,\mu+a\rho\rangle\bigr)},
    }\label{eq:norm_(anti-)symmetric_A}\\
    \fl\eqalign{
    \lefteqn{\langle H_{\mu}^{(B)\pm},H_{\nu}^{(B)\pm}\rangle_{(B)}}\\
    \Musr = & \ \delta_{\mu,\nu}
    \dfrac{N!}{\omega^{N(N-1)a+N(b+\frac{1}{2})+|\mu|}}\\
    & \prod_{j\in I}
    \Gamma\bigl(\bigl[\frac{1}{2}(\mu_j+1)\bigr]+a(N-j)+b+\frac{1}{2}\bigr)
    \Gamma\bigl(\bigl[\frac{1}{2}\mu_j\bigr]+a(N-j)+1\bigr) \\
    & \prod_{\alpha\in R_+}\frac{%
    \Gamma\bigl(\langle\alpha^\vee,\frac{1}{2}\mu 
    +a\rho\rangle +1\mp a\bigr)%
    \Gamma\bigl(\langle\alpha^\vee,\frac{1}{2}\mu
    +a\rho\rangle \pm a\bigr)}%
    {\Gamma\bigl(\langle\alpha^\vee,\frac{1}{2}\mu
    +a\rho\rangle +1\bigr)%
    \Gamma\bigl(\langle\alpha^\vee,\frac{1}{2}\mu
    +a\rho\rangle\bigr)}.
    }\label{eq:norm_(anti-)symmetric_B}
  \end{eqnarray}
  \label{eq:norm_(anti-)symmetric}
\end{subequations}
\label{th:square_norms_for_(anti-)symmetric_polynomials}
\end{theorem}
\Proof The orthogonality follows from that for the non-symmetric
case. The square norms are straightforwardly calculated from Lemma
\ref{lm:relative_norms_for_composition}, Proposition 
\ref{prop:(anti-)symmetrization} and Lemma \ref{lm:from_Poincare}.
\QED

In terms of the (anti-)symmetric eigenstates of the Hamiltonians
$\hat{\mathcal H}^{(A)\pm}(a)$ and $\hat{\mathcal H}^{(B)\pm,\pm}(a,b)$,
the above formulas are expressed by
$(\Phi_{\mu}^{(A)\pm},\Phi_{\nu}^{(A)\pm})
=\langle H_{\mu}^{(A)\pm},H_{\nu}^{(A)\pm}\rangle_{(A)}$ and
$(\Phi_{\mu}^{(B)\pm},\Phi_{\nu}^{(B)\pm})
=\langle H_{\mu}^{(B)\pm},H_{\nu}^{(B)\pm}\rangle_{(B)}$.
We remark that
the square norms of the cases $H_{\mu}^{(A)+}$, $\mu\in P_+$ and 
$H_{\mu}^{(B)+}$, $\mu\in 2P_+$ were calculated in \cite{Baker_0,%
vanDiejen_1,Kakei_1} by use of limiting procedure or shift operators,
which are different from our approach.

\section{Summary}\label{sec:summary}

We have presented the Rodrigues formulas for the monic non-symmetric 
multivariable Hermite and Laguerre polynomials that give the non-symmetric
orthogonal bases of the $A_{N-1}$- and $B_{N}$-Calogero models with
distinguishable particles. The square norms of the above non-symmetric
polynomials have been algebraically calculated by employing a 
language of a root system of a finite-dimensional simple Lie
algebra. Through symmetrization and anti-symmetrization, we have
constructed the bosonic and fermionic eigenstates of the Calogero models.
The square norms of the bosonic and fermionic eigenstates are
calculated from those of their non-symmetric counterparts with the aid
of an identity derived by the Poincar\'e polynomials.

\ack
The authors express their gratitude to Prof.~M.~Wadati for continuous
encouragement and instructive advices.
They are grateful to Dr.~Y.~Komori 
for fruitful discussions and suggestive comments.
A.N.~appreciates the Research Fellowship of the
Japan Society for the Promotion of Science for Young Scientists.
H.U.~appreciates the Grant-in-Aid for Encouragement of Young Scientists
(No.~12750250) by the Japan Society for the Promotion of Science.

\appendix
\setcounter{section}{0}
\renewcommand{\thetheorem}{\Alph{section}.\arabic{theorem}}

\section{Proof of Lemma \ref{lm:from_Poincare}}
\label{app:from_Poincare}
We present a proof of Lemma \ref{lm:from_Poincare}
in the symmetric case following our previous paper \cite{Nishino_5}.
The Poincar\'{e} polynomial is an invariant polynomial
which shows remarkable properties of the Weyl group $W$ \cite{Humphreys}. 
They are defined by
\[
 \mathcal{W}(t)
 =\sum_{w\in W}
   \prod_{\alpha\in R_{w}}t_{\alpha},
\]
where $\{t_{\alpha}|\alpha\in R\}$
are $W$-invariant indeterminates, i.e., $t_{\alpha}=t_{w(\alpha)}$
for $w\in W$.
For the Weyl group of type $A_{N-1}$,
in which all the indeterminates are equal $t_{\alpha}=t$, we have
\[
  \mathcal{W}(t)=\sum_{w\in W}t^{\ell(w)}.
\] 
In what follows, we consider only the $A_{N-1}$-case.
We denote by $\mathbb{K}$
the field of rational functions over $\mathbb{C}$
in square-roots of indeterminates $\{t\}$.
To investigate the Poincar\'{e} polynomials,
Macdonald proved the following identity \cite{Macdonald_2}:

\begin{theorem}[I.~G.~Macdonald]\hfill
\begin{equation}
 \mathcal{W}(t)=
 \sum_{w\in W}\prod_{\alpha\in R_{+}}
  \frac{1-tx^{w(\alpha^{\vee})}}{1-x^{w(\alpha^{\vee})}}.
 \label{eq:Poincare_1}
\end{equation}
\end{theorem}
\begin{lemma}
Let $\mu\in P_{+}$. We have
\begin{equation}
 \label{eq:w-sum}
 \sum_{\nu\in W(\mu)}
 \prod_{\alpha\in R_{w_{\nu}}}
   \frac{t(1-t^{-1}q^{\langle\alpha^{\vee},\mu+a\rho\rangle})}
        {1-tq^{\langle\alpha^{\vee},\mu+a\rho\rangle}}
 =\mathcal{W}(t)\prod_{\alpha\in R_{+}}
   \frac{1-q^{\langle\alpha^{\vee}\!,\mu+a\rho\rangle}}
        {1-tq^{\langle\alpha^{\vee}\!,\mu+a\rho\rangle}}.
\end{equation}
\end{lemma}

\noindent
\Proof
Define a lattice $Q^{\vee}$ by
$Q^{\vee}\Define\bigoplus_{j\in\check{I}}{\mathbb Z}\alpha^{\vee}_j$.
There exists a $\mathbb{K}$-homomorphism
$\varphi :
 \mathbb{K}[Q^{\vee}]\rightarrow \mathbb{K}$ defined by
\[
 \varphi:
 x^{\alpha_{i}^{\vee}}
 \mapsto q^{\langle\alpha_{i}^{\vee},\mu+a\rho\rangle},\,
 \text{ for }\, i\in \check{I}.
\]
Since
$\mathcal{W}(t)\in \mathbb{K}[Q^{\vee}]$ does not
depend on $\{x^{\alpha_{i}^{\vee}}\}$ as (\ref{eq:Poincare_1}), we have
\begin{eqnarray}
 \varphi\big(\mathcal{W}(t)\big) & = \mathcal{W}(t) \nn\\
 &=\sum_{w\in W}\prod_{\alpha\in R_{+}}
   \varphi\Big(
   \frac{1-tx^{w(\alpha^{\vee})}}{1-x^{w(\alpha^{\vee})}}
   \Big) \nn\\
 &=\sum_{w\in W}\prod_{\alpha\in R_{+}}
   \frac{1-tq^{\langle w(\alpha^{\vee}),\mu+a\rho\rangle}}
        {1-q^{\langle w(\alpha^{\vee}),\mu+a\rho\rangle}} \nn\\
 &=\frac{\displaystyle{
       \sum_{w\in W}
       \prod_{\alpha\in R_{w}}
       (t\!-\!q^{\langle\alpha^{\vee},\mu+a\rho\rangle})\!\!\!\!
       \prod_{\alpha\in R_{+}\setminus R_{w}}\!\!\!\!
       (1\!-\! t
        q^{\langle\alpha^{\vee},\mu+a\rho\rangle})
      }}
     {\displaystyle{
       \prod_{\alpha\in R_{+}}
       (1\!-\! q^{\langle\alpha^{\vee},\mu+a\rho\rangle})
      }} \nn\\
 &=\prod_{\alpha\in R_{+}}\!\!
   \frac{1\!-\! tq^{\langle\alpha^{\vee},\mu+a\rho\rangle})}
        {1\!-\! q^{\langle\alpha^{\vee},\mu+a\rho\rangle}}
   \sum_{w\in W}
   \prod_{\alpha\in R_{w}}\!\!
   \frac{t(1\!-\!t^{-1}
        q^{\langle\alpha^{\vee},\mu+a\rho\rangle})}
        {1\!-\!tq^{\langle\alpha^{\vee},\mu+a\rho\rangle}}. \nn
\end{eqnarray}
Thus we obtain the following relation:
\begin{eqnarray}
 \label{eq:relation}
 \sum_{w\in W}\prod_{\alpha\in R_{w}}
   \frac{t(1-t^{-1}q^{\langle\alpha^{\vee},\mu+a\rho\rangle})}
        {1-tq^{\langle\alpha^{\vee},\mu+a\rho\rangle}}
 =\mathcal{W}(t)\prod_{\alpha\in R_{+}}
   \frac{1-q^{\langle\alpha^{\vee}\!,\mu+a\rho\rangle}}
        {1-tq^{\langle\alpha^{\vee}\!,\mu+a\rho\rangle}}.
\end{eqnarray}
We show that the sum on the left-hand side of the above equation
can be replaced by the sum on $\nu\in W(\mu)$.
Consider the isotropy group $W_{\mu}=\{w\in W|w(\mu)=\mu\}$
for the partition $\mu\in P_{+}$.
Since an element $w\in W_{\mu}\setminus\{1\}$
can be written by a product of
simple reflections fixing $\mu$,
$\{s_{i}|i\in J\subset I\}$,
there exists at least one simple root $\alpha_{i}\in \Pi$
associated with the reflection $s_{i}$ in the set $R_{w}$.
Hence, for $w\in W_{\mu}\setminus\{1\}$, we have
\begin{eqnarray}
 \fl\prod_{\alpha^{\vee}\in R_{w}^{\vee}}
 t(1-t^{-1}q^{\langle\alpha^{\vee},\mu+a\rho\rangle})
 &=t(1-t^{-1}q^{\langle\alpha_{i}^{\vee},a\rho\rangle})\hspace{-5mm}
 \prod_{\alpha\in R_{w}\setminus \{\alpha_{i}\}}
  \hspace{-5mm}t(1-t^{-1}
               q^{\langle\alpha^{\vee},\mu+a\rho\rangle}) \nn\\
 &=t(1-t^{-1}t)\hspace{-5mm}
  \prod_{\alpha\in R_{w}\setminus \{\alpha_{i}\}}
  \hspace{-5mm}t(1-t^{-1}
               q^{\langle\alpha^{\vee},\mu+a\rho\rangle})=0. \nn
\end{eqnarray}
Define $W^{\mu}\Define \{w\in W|\ell(ws_{i})>\ell(w)\,
\text{ for all }\, i\in J\}$.
For $w\in W$, there is a unique $u\in W^{\mu}$
and a unique $v\in W_{\mu}$ such that $w=uv$.
The formula $R_{w}=R_{v}\cup v^{-1}R_{u}$ shows that,
if $v\neq 1$, the product on $\alpha\in R_{w}$ 
in (\ref{eq:relation}) vanishes.
Thus we obtain the above lemma since the sum on $w\in W$ on
the left-hand side of (\ref{eq:relation})
can be replaced by that on $u\in W^{\mu}$
which is equivalent to that on $\nu\in W(\mu)$.
\QED

In the formal limit $q\rightarrow 1$ under the restriction
$t=q^{a}$, we have the relation in Lemma \ref{lm:from_Poincare}.
The formula in the anti-symmetric case is proved in a similar way.


\end{document}